\documentclass[11pt]{article}
\usepackage{amsmath,amssymb,amsthm,latexsym}
\usepackage{dsfont}
\usepackage[english]{babel}
\usepackage{graphicx,color}

\usepackage[top=2cm, bottom=2cm, left=2cm, right=2cm]{geometry}
\usepackage{amsfonts}

\usepackage{braket}
\newcommand{\bea}{\begin{eqnarray}}
\newcommand{\eea}{\end{eqnarray}}
\newcommand{\dr}{\partial}
\newcommand{\tr}{\mathrm{Tr}}
\newcommand{\lnz}{\mathrm{ln} Z}
\newcommand{\res}{\mathcal R}
\newcommand{\mb}{\mathbf}

\usepackage{amscd}

\usepackage[pdftex]{hyperref}
\newtheorem{lemma}{Lemma}

\newtheorem{definition}{Definition}
\newtheorem{theorem}{Theorem}
\newtheorem{corollary}{Corollary}

\newcommand{\cD}{{\cal D}}
\newcommand{\cT}{{\cal T}}
\newcommand{\cC}{{\cal C}}
\newcommand{\cL}{{\cal L}}

\allowdisplaybreaks[4]

\begin{document}

\title{Universality and Borel Summability\\ of Arbitrary Quartic Tensor Models}
 
\author{
Thibault Delepouve\footnote{delepouve@cpht.polytechnique.fr, Laboratoire de Physique Th\'eorique, CNRS UMR 8627, Universit\'e Paris Sud, 91405 Orsay Cedex, France
 and Centre de Physique Th\'eorique, CNRS UMR 7644, \'Ecole Polytechnique, 91128 Palaiseau Cedex, France.},
 \ Razvan Gurau\footnote{rgurau@cpht.polytechnique.fr, Centre de Physique Th\'eorique, CNRS UMR 7644, \'Ecole Polytechnique, 91128 Palaiseau Cedex, France
 and Perimeter Institute for Theoretical Physics, 31 Caroline St. N, N2L 2Y5, Waterloo, ON, Canada.} \ and
Vincent Rivasseau\footnote{rivass@th.u-psud.fr, Laboratoire de Physique Th\'eorique, CNRS UMR 8627, Universit\'e Paris Sud, 91405 Orsay Cedex, France
 and Perimeter Institute for Theoretical Physics, 31 Caroline St. N, N2L 2Y5, Waterloo, ON, Canada.}}

\maketitle

\begin{abstract}
We extend the 
study of \emph{melonic} quartic tensor models to models with arbitrary quartic interactions.
This extension requires a new version of the loop vertex expansion using several species of intermediate fields and 
iterated Cauchy-Schwarz inequalities. Borel summability is proven, uniformly as the tensor size $N$ becomes large.
Every cumulant is written as a sum of explicitly calculated terms plus a remainder, suppressed in $1/N$.
Together with the existence of the large $N$ limit of the second cumulant,
this proves that the corresponding sequence of probability measures  
is uniformly bounded and obeys the tensorial universality theorem.

\end{abstract}

\begin{flushright}
\end{flushright}
\medskip

\section{Introduction}

Matrix models \cite{Mehta,matrix} allow to study in two dimensions critical phenomena 
on a random geometry
and the quantization of gravity coupled with conformal matter.
Tensor models \cite{review} are their generalization to higher dimensions. 
Colored \cite{color} and
invariant \cite{uncoloring} tensor models 
support a $1/N$ expansion
\cite{expansion1,expansion2,expansion3}. This analytical tool is crucial for establishing their large $N$ 
and double scaling limit \cite{largeN,dtens1,dtens2} as well as their application 
to the study of critical phenomena in random geometries \cite{b01,b02,b1,b2,b3,melo1}
and possibly to quantization of gravity \cite{tt3} in higher dimensions.

Already for quartic interactions, tensor models have a very rich structure: whereas there exists
a unique quartic invariant one can build out of a matrix, $\tr\left[ MM^{*}MM^{*}\right]$, there are numerous 
possible quartic invariants for tensors, as the indices of four tensors can be contracted in many different ways.

The mathematical study of tensor models as probability measures 
has recently been started \cite{universality},
with the rigorous non-perturbative construction of a specific quartic tensor model completed in 
\cite{RGurau} using 
the Loop Vertex Expansion (LVE) \cite{LVE1,LVE2,MagRiv}. While this model is symmetric over all the colors,
hence is a genuine tensor model, and not just a model for a rectangular matrix, it includes only the simplest 
possible quartic interactions, called \emph{melonic}. In particular, for this melonic model,
a factorization property holds which is crucial for establishing the results in \cite{RGurau}.

The factorization fails as soon as one considers a slightly more general model:  even adding just some non melonic 
quartic invariants spoils it. This brings the natural question: are the results of \cite{RGurau} generic, or are they the result of
a lucky accident? In this paper we prove that these non perturbative results and the tensorial universality theorem of \cite{universality}  
hold for any quartic tensor model. However, 
as the factorization fails, we need to use an entirely different technique to establish them.

This paper is organized as follows. Section \ref{sec:prereq} introduces the framework of tensor models with an arbitrary quartic interactions
and presents our theorems. The rest of the sections contains the proofs.

\section{The model and the main results}\label{sec:prereq}

In this section we introduce the notations and state our main results. The proofs of these results are
presented starting with the next section. 

\subsection{Generalities}

Let us consider a Hermitian inner product space $V$ of dimension $N$ and $\{e_n| n=1,\dots N\}$ an orthonormal basis in $V$. 
The dual of $V$, $V^{\vee}$ is identified with the complex conjugate $\bar V$ via the conjugate linear isomorphism 
\[
z \to  z^{\vee}(\cdot)=\langle z, \cdot\rangle  \; . 
\]
We denote $e^n \equiv e_n^{\vee} = \langle e_n , \cdot \rangle$ the basis dual to $e_n $. Then
\[
 \Bigl(\sum_n z^n e_n  \Bigr)^{\vee}(\cdot) =  \sum_n \overline{ z^n }\langle e_n, \cdot\rangle = \sum_{n} (z^{\vee })_n e^n (\cdot)
 \Rightarrow  (z^{\vee })_n = \overline{ z^n } \; .
\]
A covariant tensor of rank $D$ is a multilinear form ${\bf T}:  V^{\otimes D} \to \mathbb{C}$. 
We denote its components in the tensor product basis by 
\[ 
T_{n^1\dots n^D} \equiv {\bf T} (e_{n^1},\dots, e_{n^D}) \; , \qquad {\bf T} = \sum_{n^1,\dots n^D} T_{n^1\dots n^D} \; \;e^{n^1} \otimes \dots \otimes e^{n^D} \; .
\]
A priori $T_{n^1\dots n^D}$ has no symmetry properties, hence its indices have a well defined position. 
We call the position of an index its \emph{color}, and we denote $\cD$ the set of colors $\{1,\dots D\}$. 

A tensor can be seen as a multilinear map between vector spaces. There are in fact as many choices as there are subsets $\cC\subset \cD$:
for any such subset the tensor is a multilinear map ${\bf T}: V^{\otimes \cC} \to \bar V^{ \otimes \cD \setminus \cC}$:
\[
 {\bf T}(z^{(c)}, c \in  \cC ) = \sum_{n^c, c\in  \cC} T_{n^1\dots n^D} \prod_{c\in   \cC} [z^{(c)}]^{n^c} \; .
\]
We denote $n^{\cC}= (n^c,c\in \cC)$ the indices with colors in $\cC$.
The complementary indices are then denoted $n^{\cD\setminus \cC} =( n^c,c\notin \cC)$. In this notation the set of all the indices 
of the tensor should be denoted $n^{\cD}=(n^1,\dots n^D)$. We will use whenever possible the shorthand notation $n\equiv n^\cD$.
The matrix elements of the linear map (in the appropriate tensor product basis) are
\[
  T_{ n^{\cD\setminus \cC} n^{\cC} } \equiv T_n \equiv T_{n^1\dots n^D} \; \text{ with } n^c \in n^{\cC} \cup n^{\cD \setminus \cC} \;,\;\; \forall c \; .
\]

As we deal with complex inner product spaces, the dual tensor $ {\bf T}^{\vee}$ is defined by 
\[ {\bf T}^{\vee} \left( (z^{(1)})^{\vee}, \dots  (z^{(D)})^{\vee} \right) \equiv
  \overline{ {\bf T} \left( z^{(1)}, \dots  z^{(D)} \right) } \; .
\]
Taking into account that 
\[
   \sum_{n^{\cD}} \overline{T_{n^1\dots n^D}} \;  \overline{ (z^{(1)} )^{n^1} } \dots \overline{ (z^{(D)} )^{n^D} }
= \sum_{n^{\cD}} \overline{T_{n^1\dots n^D}} \;   ( z^{(1)\vee})_{n^1}  \dots  (z^{(D)\vee})_{n^D} \;,
\]
we obtain the following expressions for the  dual tensor and its components
\[  {\bf T}^{\vee} = \sum_{n^1, \dots n^D} \overline{T_{n_1\dots n_D}}\; \; e_{n^1} \otimes \dots e_{n^D} \; , \;\;
 ({\bf T}^{\vee})^{n^1\dots n^D} =  \overline{T_{n_1\dots n_D}} \; .
\]
The dual tensor is a conjugated multilinear map 
${\bf T}^{\vee} :  \bar V^{\otimes \cD \setminus\cC} \to   V^{\otimes \cC}  $
with matrix elements 
\[
 ( T^{\vee})^{ n^{\cC}  n^{\cD \setminus \cC} } \equiv  \overline{ T_{ n^1\dots  n^D} } \; \text{ with } 
  n^c \in n^{\cC} \cup  n^{\cD \setminus \cC} \; ,\;\; \forall c \; .
\]
From now on we denote $ \bar T_{n_1\dots n_D} \equiv \overline{T_{n_1\dots n_D}} $, we write all the indices in subscript,
and we denote the contravariant indices with a bar.
Indices are always understood to be listed in increasing order of their colors.
We denote $ \delta_{n^{\cC} \bar n^{\cC}} = \prod_{c\in \cC} \delta_{n^c \bar n^c}  $ and 
$\tr_{\cC}$ the partial trace over the indices $n^c, c\in \cC$.

\subsection{Trace invariants and tensor models}

Under unitary base change, covariant tensors transform under the tensor product of $D$ fundamental 
representations of $U(N)$: the group acts independantly on each index of the tensor. 
For $U^{(1)}...U^{(D)}\in U(N)$,
\[
 {\bf T} \to  \left( U^{(1)} \otimes ... \otimes U^{(D)} \right){\bf T} , 
 \qquad  {\bf T}^{\vee} \to {\bf T}^{\vee} \left( U^{(1)*} \otimes ... \otimes U^{(D)*} \right). 
\]
In components, it writes,
\[
 T_{a^{\cD}}\to \sum_{m^{\cD}} U^{(1)}_{a^1 m^1}...U^{(D)}_{a^D m^D}\ T_{m^{\cD}},
 \qquad  \bar{T}_{\bar a^{\cD}}\to \sum_{m^{\cD}}  \bar U^{(1)}_{\bar a^1 \bar m^1}... \bar U^{(D)}_{\bar a^D \bar m^D}\ \bar T_{\bar m^{\cD}}.
\]

 A \emph{trace invariant} is a invariant quantity under the action of the external tensor product of $D$ 
 independant copies of the unitary group $U(N)$  which is built by contracting indices of a product of tensor entries.

The tensor and its dual can be composed as linear maps to yield a map from $V^{\otimes \cC}$ to $V^{\otimes \cC}$
\[
 [ {\bf T}^{\vee} \cdot_{\cD\setminus \cC }  {\bf T} ]_{\bar n^{  \cC} n^{  \cC}  }
 = \sum_{n^{\cD\setminus\cC} , \bar n^{\cD \setminus \cC} }
  \bar T_{ \bar n^1\dots \bar n^D}
  \delta_{\bar n^{\cD \setminus \cC}  n^{\cD \setminus \cC}   }  T_{ n^1\dots n^D}   \; .
\]

The unique quadratic trace invariant is the (scalar) Hermitian pairing of ${\bf T}^{\vee}$ and ${\bf T}$ which 
writes:
\[
 {\bf T}^{\vee} \cdot_{\cD  }  {\bf T} = \sum_{n^{\cD} \bar n^{\cD}}
  \bar T_{ \bar n^1\dots \bar n^D}  \delta_{ \bar n^{\cD} n^{\cD}}  T_{ n^1\dots n^D}   \; ,
\]

A connected quartic trace invariant $V_{\mathcal{C}}$ for ${\bf T}$ is
specified by a subset of indices $\cC \subset \cD $:
\[
 V_{\cC}({\bf T}^{\vee},{\bf T} ) = \tr_{\cC} \Big[ \left[ {\bf T}^{\vee} \cdot_{\cD\setminus \cC }  {\bf T} \right] \cdot_{\cC}
 \left[{\bf T}^{\vee} \cdot_{\cD\setminus \cC }  {\bf T} \right]  \Big] \;,
\]
where we denoted $\cdot_{\cC}$ the product of operators from $V^{\otimes \cC}$ to $V^{\otimes \cC}$.
In components this invariant writes:
\[
 \sum_{n, \bar n, m, \bar m} 
  \left(  \bar{T}_{\bar{n}}\  \delta_{\bar n^{\cD\setminus \cC} n^{\cD\setminus \cC}  }  \  T_{n}    \right) \ 
   \delta_{n^{\cC}\bar m^{\cC}} \delta_{   \bar n^{\cC}  m^{\cC}}\ 
   \left( \bar{T}_{\bar{m}}\  \delta_{\bar m^{\cD\setminus \cC}  m^{\cD\setminus \cC}  }  \ T_{m}    \right)   \; .
\]
 
A generic quartic tensor model is then the (invariant) perturbed Gaussian measure for a random tensor:
\begin{align}\label{eq:model}
d\mu = \Bigl( \prod_n N^{D-1} \frac{ d \bar{T}_{{n}}  dT_n }{2 \imath \pi} \Bigr) \,
 e^{ -N^{D-1} \big(  {\bf T}^{\vee} \cdot_{\cD  }  {\bf T} 
  +\lambda\sum_{\mathcal{C} \in \mathcal{Q}} V_{\mathcal{C}}({\bf T}^{\vee},{ \bf T} ) \big) } \;,
\end{align}
where  $\mathcal{Q}$ is some set of $\cC$s. From now, we will denote $\mathcal N_\mathcal{Q} = |\mathcal{Q}|$, the cardinal of $\mathcal{Q}$.

The melonic models previously treated in the literature \cite{RGurau} are obtained by  
restricting to $\mathcal{Q}=\{\mathcal{C},|\mathcal{C}|=1 \}$. Considering arbitrary $\mathcal{Q}$s
has important consequences. One of the features of the melonic model of \cite{RGurau} is that, in the loop vertex expansion, 
the amplitude of graphs factors in contributions associated to the faces (technically this is 
done by introducing Schwinger parameters on the resolvents) and one immediately recovers the appropriate scaling with $N$. 
This does not hold in the general model. Recovering the appropriate scaling in $N$ requires a new technique in this more
general case relying on iterated Cauchy-Schwarz inequalities which we will present below.
\medskip

The moment-generating function of the measure $d\mu$ is defined as :
\begin{align*}
& Z(J, \bar{J})=\int d\mu \;\; e^{ \sum_{n} T_{n} \bar{J}_{n} + \sum_{\bar n}\bar{T}_{\bar{n} }J_{\bar{n} }} \;,
\end{align*}
and its cumulants are thus written :
\begin{align*}
& \kappa(T_{n_1}\bar{T}_{\bar{n}_1}...T_{n_k}\bar{T}_{\bar{n}_k})
=\frac{\partial^{(2k)} \Bigl( \ln Z(J,\bar J) \Bigr) }{\partial \bar{J}_{n_1}
\partial J_{\bar{n}_1}...\partial \bar{J}_{n_k}\partial J_{\bar{n}_k}} \Bigg{\vert}_{J =\bar J =0}.
\end{align*}

\subsection{Gaussian measure and universality}

The Gaussian measure of covariance $\sigma^2$ for a random tensor is:
\begin{align*}
d\mu_{G} = \Bigl( \prod_n \frac{N^{D-1}}{\sigma^2} \frac{ d \bar{T}_{{n}}  dT_n }{2 \imath \pi} \Bigr) \,
 e^{ -\sigma^{-2}N^{D-1} \   {\bf T}^{\vee} \cdot_{\cD  }  {\bf T} }
 \;,
\end{align*}
For any trace invariant $\mathrm{B}({\bf T}^{\vee},{ \bf T})$ made of $k$ covariant and $k$ contravariant tensors, there are two non-negative integers, $\Omega(\mathrm{B})$ ans $R(\mathrm{B})$, 
such that  the large $N$ limit of the Gaussian expectation follows \cite{universality}
\begin{align*}
\lim_{N\to\infty}N^{\Omega(\mathrm B)-1}\mu_G \left(\mathrm B({\bf T}^{\vee},{ \bf T})\right)=R(\mathrm B)\ \sigma^{2k}.
\end{align*}
$\Omega(\mathrm B)$ is called the convergence order of B.
\begin{definition}
 A random tensor ${ \bf T}$ with the probability measure $\mu$ {\bf converges in distribution} to the distributional limit of a Gaussian tensor of covariance $\sigma^2$ if the large $N$ 
 limit of the expectation of any trace invariant $\mathrm B$ equals the limit of the Gaussian expectation of the invariant.
 \begin{align*}
  \lim_{N\to\infty}N^{\Omega(\mathrm B)-1}\mu \left(\mathrm B({\bf T}^{\vee},{ \bf T})\right)=R(\mathrm B)\ \sigma^{2k}.
 \end{align*}
\end{definition}

\begin{definition}
A random tensor ${\bf T}$ with the probability measure $\mu$ is {\bf trace invariant} if its cumulants are linear combinaisons of trace invariant operators.
\begin{align*}
\kappa(T_{n_1}\bar{T}_{\bar{n}_1}...T_{n_k}\bar{T}_{\bar{n}_k})
=\sum_{\pi, \bar\pi}\sum_{\rho_{\cD} }\mathfrak{K}(\rho_{\cD})\prod_{d=1}^k  
\prod_{c=1}^D\delta_{n_{\pi(d)}^c \bar{n}_{\rho_c\bar\pi(d)}^c} \;,
\end{align*}
where $\rho_{\cD}=(\rho_1, \dots \rho_D)$ runs over $D$-uples of permutations of $k$ elements, 
$\pi$ and $\bar\pi$ are permutations over $k$ elements, and $\mathfrak{K}(\rho_{\cD})$ depends only on $\rho_{\cD}$, $\lambda$ and $N$. 
\end{definition}

We denote $\mathfrak{C}(\rho_{\cD})$ the number of connected components of the graph associated to $\rho_{\cD}$. 
(We represent $\rho_\cD$ as a $D$-colored bipartite graph with $k$ black and $k$ white $D$-valent vertices, each set being indexed form $1$ to $k$, 
and the white vertex $l$ is connected to the black vertex $\rho_c (l)$ by an edge of color $c$. Then, $\mathfrak{C}(\rho_{\cD})$ is the number of connected components of this graph).

\begin{definition}
 A trace invariant probability measure $\mu$ is {\bf properly uniformly bounded} at large $N$ if
 \begin{align*}
  \lim_{n\to\infty} N^{D-1} \mathfrak{K}(\{1\}_{\mathcal{D}}) < \infty,
 \end{align*}
 with $\{1\}_{\mathcal{D}}$ being the $D$-uples of trivial permutations over $k=1$ element, and if
 \begin{align*}
  N^{ 2(D-1) k  - D +\mathfrak{C}(\rho_{\cD})  } \mathfrak{K}(\rho_{\mathcal{D}}) < K(\rho_{\mathcal{D}}),\ \forall k\neq 1,\ \forall N, 
 \end{align*}
 for some constant $K(\rho_{\mathcal{D}})$.
\end{definition}

The main result of this paper is to prove that for a perturbation parameter $\lambda\in [\, 0, \frac{1}{8\mathcal N_{\mathcal{Q}}})$, quartic tensor models are
properly bounded trace invariant probability distributions, and thus obey the second universality theorem in \cite{universality}.

\medskip

{\bf Theorem.} (Universality)
\emph{ For large $N$, a random  tensor whose probability distribution $\mu$ is trace invariant and properly uniformly bounded converges in distribution to a Gaussian tensor of covariance
$\sigma^2=\lim_{N\to\infty}\ N^{D-1} \mathfrak{K}(\,\{1\}_{\mathcal{D}}\,)$.}

\subsection{BKAR Formula}
In order to compute the cumulants of $\mu$ we need to compute the logarithm of the generating function $Z(J,\bar J)$.
This can be done by using at first a replica trick, then 
the Brydges-Kennedy-Abdesselam-Rivasseau (BKAR) forest formula \cite{BK,AR,garden} interpolating the Gaussian measure between the replicas, 
and finally expressing the logarithm as a sum over trees.

Let $X$ be a complex vector, for any function $V(X,\bar X)$, denoting $d\mu_C$  the Gaussian measure of covariance $C$,
\[
 \int d\mu_C \;\; X_a \bar X_b = C_{ab} \; ,
\]
we have:
\begin{align}\label{eq:forest}
& \mathrm{ln}\int e^{V(X,\bar X)}d\mu \quad =\quad  
 \sum_{v\geq 1}\frac1{v!} \sum_{T_v}\int_0^1 \left(\prod_{l_{ij}\in T_v}du_{ij}\right)\int d\mu_{T_v,u}\nonumber\\
&\times\quad \left[\prod_{l_{ij}\in T_v}\left(
\sum_{a,b}\left(
\frac{\dr}{\dr X_{a}^{i}}C_{ab}\frac{\dr}{\dr\bar{X}_{b}^{j}} + \frac{\dr}{\dr X_{a}^{j}}C_{ab}\frac{\dr}{\dr\bar{X}_{b}^{i}}
\right)\right)
\right]\prod_{i=1}^v V\left(X^i,\bar X^i\right),
\end{align}
where $T_v$ runs over combinatorial trees with $v$ vertices labelled from $1$ to $v$,
$l_{ij}$ denotes the edge of the tree connecting the vertices $i$ and $j$ and $u_{ij}$ is an interpolation parameter running from $0$ to $1$.
The $X^i$ and $\bar X^i$ are random complex vectors associated with the vertices $i$ of the tree and distributed with 
the interpolated Gaussian measure $d\mu_{T_v,u}$ defined as:
\begin{align*}
 \int d\mu_{T_v, u} \;\;\; X^i_{a}\bar{X}^j_{b}= w_{ij}C_{ab} \;,
\end{align*}
where, denoting $\mathcal P_{ij}$ is the unique path in $T_v$ joining the vertices $i$ and $j$, 
\begin{equation}\label{eq:wdef}
 w_{ij} = \begin{cases}
           1 \qquad & i=j \\
           \min_{ l_{km}\in \mathcal P_{ij}  }\{u_{km}\} \quad & i\neq j
          \end{cases} \; .
\end{equation}
The matrix $w_{ij}$ is positive, hence the measure $d\mu_{T_v,u} $ is well defined.

Remark that the integral with measure $  d\mu_{T_v, u}$ can be rewritten as:
\begin{align*}
& \int d\mu_{T_v, u} \; \; F(X,\bar X) =\crcr
& \quad = \Big[e^{  \sum_{i,j} w_{ij} \Bigl[ \sum_{ab} C_{ab} 
\left( \frac{\dr} {\dr X_{a}^{i} }  \frac{\dr}{\dr\bar X_{b}^{j} } \right) \Bigr] } F(X, \bar X)\Big]_{X=0} \; ,
\end{align*}
where the matrix $w_{ij}$ is symmetric, namely $w_{ij} = w_{ji}$.

\subsection{Summary of the results}
We denote $\mathcal N_{\mathcal{Q}}=|\mathcal{Q}|$ the number of distinct interaction terms in a given model (hence for melonic 
models $\mathcal N_{\mathcal{Q}}=D$). Note that $\cC$ and $\cD\setminus \cC$ play a complementary role in a trace invariant. 
Given an invariant, there are two natural choices of $\cC$: 
\begin{itemize}
 \item $\cC$ are the colors shared by two tensors such that the color $1$ never belongs to $\cC$, $1\notin \cC$.
 \item $\cC$ are the colors shared by two tensors such that $|\cC|\le |\cD \setminus \cC|$, that is $|\cC|\le D/2$.
\end{itemize}

\medskip

\noindent{\bf The Loop Vertex Expansion.} A first set of results of this paper concerns the loop vertex expansion 
(introduced in \cite{LVE1,LVE2}) of the cumulants of the measure in eq. \eqref{eq:model}. 

Let us denote $\sigma^{\cal{C}}_{a^{\cal C} b^{\cal C}}$ a $N^{|\cal{C}|} \times N^{|\cal{C}|}$ matrix with  
line and column indices of colors in $\cal{C}$, $a^{\cal {C}} = (a^c | c\in \cal{C})$, and let us denote 
$\mathbf1^{\cD\setminus \cC}$ the identity matrix on the indices of colors $\cD\setminus \cC$. 
We define  
\begin{align*}
& A(\sigma)=\sqrt{\frac{\lambda}{ N^{D-1}}}\left(  \sum_{\mathcal{C}\in \cal{Q}}\mathbf1^{\cD \setminus \cC }
\otimes(\sigma^{\mathcal{C}}-\sigma^{{\mathcal{C}}*}) \right) \; ,\crcr
& \big[A(\sigma) \big]_{nm} = \sqrt{\frac{\lambda}{ N^{D-1}}}  
\sum_{\mathcal{C}\in \cal{Q}} \delta_{n^{\cD\setminus \cC} m^{\cD\setminus \cC}} \;\; 
(\sigma^{\mathcal{C}}-\sigma^{{\mathcal{C}}*})_{n^{\cal{C}} m^{\cal{C}} }  \; .
\end{align*}
 Note that, as $A(\sigma)$ is anti Hermitian, $ [ \mathbf1^{\cD} + A(\sigma) ]^{-1}$ is 
well defined for all $\lambda\in \mathbb{C} \setminus (-\infty,0)$.

The loop vertex expansion of the generating function of the moments of $\mu$ is:
\begin{lemma}\label{lem:lemmaLVE}
The generating function of the moments of $\mu$ (i.e. the partition function of the quartic model) with a set of 
interactions $\mathcal{Q}$ is:
\begin{align}\label{eq:ZLVE}
&Z(J,\bar J)=\int  \prod_{\mathcal{C} \in \mathcal{Q}}\prod_{a^{\cal {C}}b^{\cal {C}}} 
\frac{d\sigma_{a^{\cal {C}}b^{\cal {C}}}^{\mathcal{C}}d\bar{\sigma}_{a^{\cal {C}}b^{\cal {C}}}^{\mathcal{C}}}{2\imath\pi} 
\;\; e^{-\ \sum_{ \mathcal{C} \in \cal{Q} }\tr_{\cC} \big[\sigma^{{\mathcal{C}}*}\sigma^{\mathcal{C}}\big] \ } 
\crcr
& \qquad \times e^{-
\ \tr_{\cD} \big[ \ln\left(\mathbf1^{\cD} + A(\sigma)\right)   \big] 
+ \frac{1}{N^{D-1}} \sum_{nm } \bar J_{n} 
\left[\frac{1}{ \mathbf1^{\cD} + A(\sigma) }\right]_{nm}J_{m} }  \; .
\end{align}
\end{lemma}

This lemma is proven in subsection \ref{subsecIFR}. 

The next theorem establishes 
the loop vertex expansion of the cumulants of $\mu$. This expansion relies on the BKAR formula
adapted to tensor models, which requires adding a number of twists.

The cumulants of $\mu$ are expressed as sums over \emph{plane trees with marked vertices and colored edges}. Plane trees have a 
well defined ordering at the vertices, and a mark is a specified starting point for this ordering. 
The edges of the trees are colored by subsets of colors ${\cal C}$. We have here the first important difference
between the melonic model of \cite{RGurau} and the general case presented here: in the former case the edges of the trees had
a unique color, while now they can carry several.

Let us denote $\mathcal{T}_{v,\{i_d\},\{{\mathcal{C}}(l)\}}$ a plane tree with $v$ vertices, labelled $1,2\dots v$,
and whose vertices $\{i_d\}, d=1\dots k$ are marked. We denote ${\mathcal{C}}(l_{ij})$ the colors of the tree edge $l_{ij}$, 
and $T_v$ the combinatorial tree associated to $\mathcal{T}_{v,\{i_d\},\{{\mathcal{C}}(l)\}}$.

The $\sigma^{\cal C},\bar \sigma^{\cal C}$ fields are now replicated over the vertices, $\sigma^{{\cal C}\ i}$ and $\bar \sigma^{{\cal C} \ i}$ 
and the interpolated Gaussian measure $d\mu_{T_v,u}(\sigma)$ is degenerated over the colorings ${\cal C}$:
\[
  e^{\sum_{ij}w_{ij} \left[ \sum_{\mathcal{C}; a^{\cal C} b^{\cal C} }\left(
\frac{\dr}{\dr\sigma_{a^{\cal C}b^{\cal C}}^{ {\mathcal{C} \ i  } } }   
\frac{\dr}{\dr\bar\sigma_{a^{\cal C}b^{\cal C}}^{{\mathcal{C} \ j }}}
\right)  \right]  } \; .
\]

The contribution of each tree is a certain contraction of \emph{resolvent} operators and external source terms $J$ and $\bar J$.
The resolvents are defined as:
\begin{align}\label{eq:res}
     R(\sigma)= \left[ {\bf 1}^{\cD} + A(\sigma) \right]^{-1} \; .
\end{align}
We adopt the following graphical representation. We represent every vertex  of the plane tree as a fat vertex, 
having $D$ interior strands, corresponding to the $D$ indices of the resolvent. The strands are labelled $1$ to $D$ 
from the most interior strand to the exterior one.

\begin{figure}[ht]
\centering
\begin{tabular}{ccc}
\includegraphics[height=1cm]{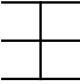} & \hspace{.5cm} & \includegraphics[height=1cm]{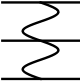} \end{tabular}
\caption{Graphical representation of a resolvent (left) and a $J_m\bar J_n$ term (right) for rank $D=3$ tensors.}
\label{figconv}
\end{figure}

Plane trees have a well defined notion of \emph{corners} which are pieces of the 
vertices comprised between two consecutive halflines. Every corner of the vertex is the represented as $D$ parallel strands crossed by a vertical line, 
as in Figure \ref{figconv} on the left. 

\begin{figure}[ht]
\centering
\includegraphics[height=5cm]{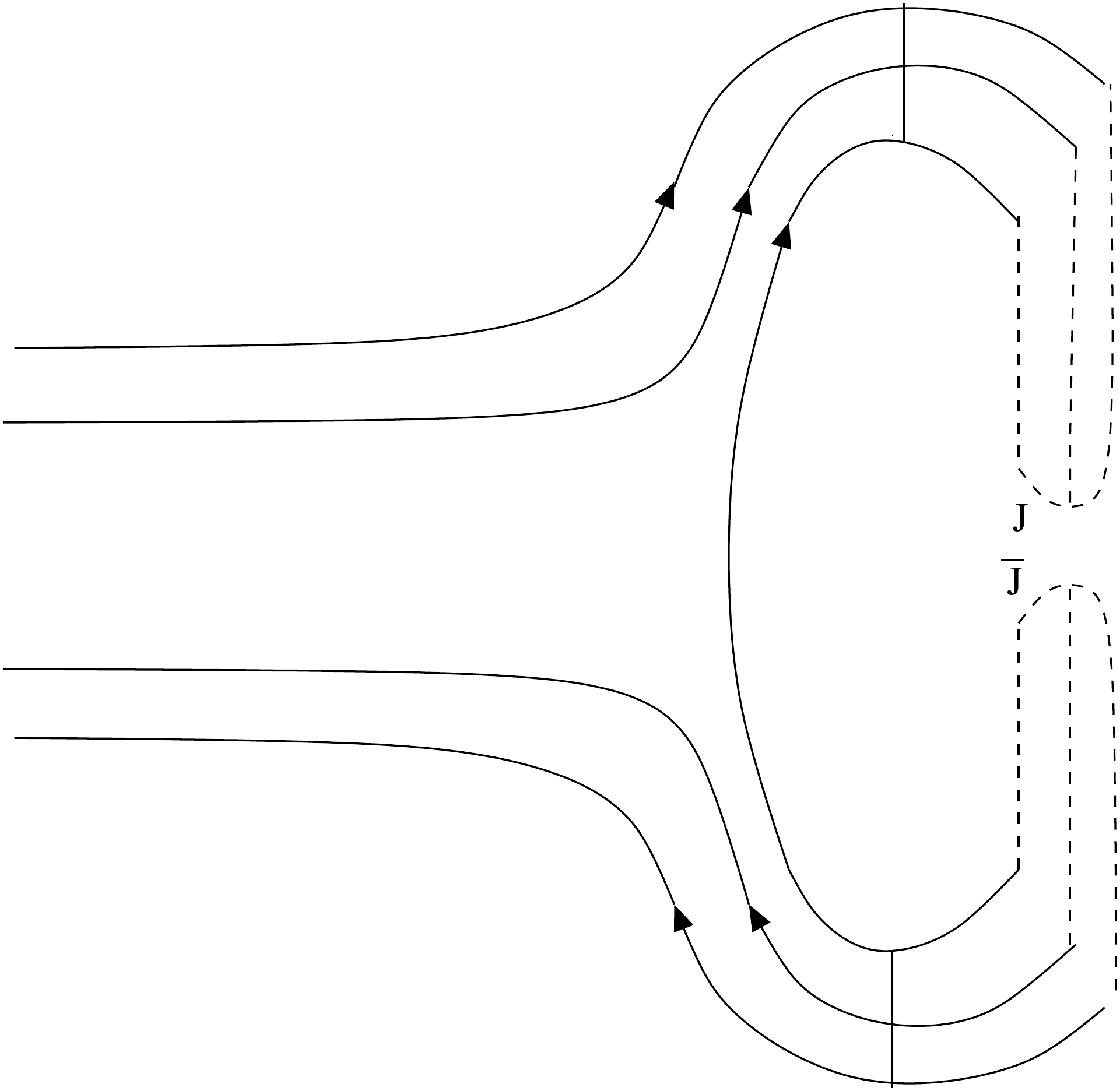}
\caption{Detailed representation of the external sources $J$ and $\bar J$.}
\label{fig:graph123}
\end{figure}

The marks correspond to $J$ and $\bar J$ sources. We either represent them as $D$ parallel strands crossed by with 
a wiggly line (as in Figure \ref{figconv} on the right) or as a pair of caps gluing together the $D$ strands 
as in Figure \ref{fig:graph123}. The caps (pictured as dashed in Figure \ref{fig:graph123}) 
represent a $J$ and a $\bar J$ source. Both 
$J$ and $\bar J$ have $D$ indices which are contracted with resolvent indices. This is pictured 
by the fact that the dashed strands hook to solid strands in Figure  \ref{fig:graph123}.

\begin{figure}[ht]
\centering
\includegraphics[height=3cm]{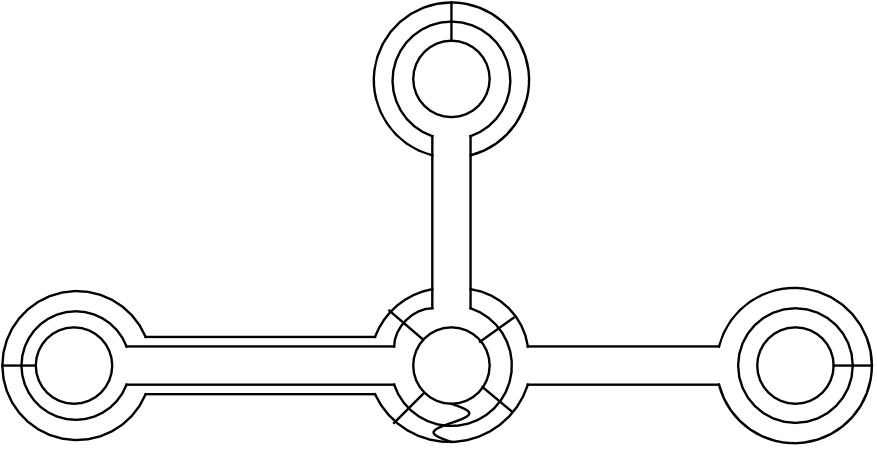}
\caption{An example of plane tree with 1 marked and 3 regular vertices, and edges of color 
$\cC = \{3  \}$, $\{2\}$ and $\{2,3\}$. Resolvents and $J\bar J$ marks are represented as in Fig.\ref{figconv} and stuck to their respective vertices.}
\label{graph1}
\end{figure}

A plane tree is then a set of edges connecting such vertices. Every edge will transmit the strands corresponding to the indices $\cC$ from one vertex to the other,
and will connect on the same vertex the indices in $\cD \setminus \cC$. Hence edges have multiple strands, as in Figure \ref{graph1}.

The number of corners of the vertex $i$, denoted $res(i)$,
is equal to the degree of the vertex if it is not marked and it is equal to the degree of the vertex plus one if it is. 
We label the corners of the vertex $i$ by $p$ in the order they are encountered when 
turning clockwise around the vertex. To the $p$'th corner of the vertex $i$ we associate a resolvent
$R(\sigma^i)_{n_{i,p} m_{i,p}}$.

To every edge of a tree we associate a contraction of the indices of the four resolvents corresponding to the four corners 
incident to the edge. If the edge $l_{ij}$ is incident to the corners $q$ and $q+1$ of the vertex $i$, and $p$ and $p+1$ of the vertex $j$,
the contraction associated to the edge is:
\begin{equation}
\label{eq:delta}
 \delta_{ \cD  }^{l_{ij},{\mathcal{C}}(l_{ij})}=  
   \left( \delta_{m^{\cD \setminus \cC }_{i,q} n^{\cD \setminus \cC }_{i,q+1} } \right) 
    \delta_{ m^{\cC}_{i,q} n^{\cC}_{j,p+1} }      \delta_{ m^{\cC}_{j,p} n^{\cC}_{i,q+1} }
 \left( \delta_{m^{\cD  \setminus \cC }_{j,p} n^{\cD  \setminus \cC }_{j,p+1} } \right) \; .
\end{equation}
 
\begin{theorem}\label{th:LVE}
The measure $\mu$ is trace invariant, and its cumulants are given by:
\begin{align}\label{eq:cumulantfin}
\kappa(T_{n_1}\bar{T}_{\bar{n}_1}...T_{n_k}\bar{T}_{\bar{n}_k})
=\sum_{\pi, \bar\pi}\sum_{\rho_{\cD} }\mathfrak{K}(\rho_{\cD})\prod_{d=1}^k  
\prod_{c=1}^D\delta_{n_{\pi(d)}^c \bar{n}_{\rho_c\bar\pi(d)}^c} \;,
\end{align}
where $\rho_{\cD}=(\rho_1, \dots \rho_D)$ runs over $D$-uples of permutations of $k$ elements, 
$\pi$ and $\bar\pi$ are permutations over $k$ elements. 
With the notation introduced above, $\mathfrak{K}(\rho_{\cD})$ is:  
\begin{align}\label{cumulants0}
 \mathfrak{K}(\rho_{\cD})  = &
\sum_{v\geq k}\frac1{v!}\ \frac{(-2\lambda)^{v-1}}{N^{(D-1)(k+v-1) }} \sum_{\tau_{\cD}} \left( \prod_{c=1}^{D} 
\mathrm{Wg}(N,\tau_c \rho_c^{-1})\right)\crcr
&\quad \times \sum_{\mathcal{T}_{v,\{i_d\},\{{\mathcal{C}}(l)\}}}
\int_0^1 \left(\prod_{l_{ij}\in T_v}du_{ij}\right)\int d\mu_{T_v,u}(\sigma)  \sum_{n,m}\\
&\quad \times\left(\prod_{i=1}^v\prod_{p=1}^{\mathrm{res}(i)}  R(\sigma^i)_{ n_{i,p}  m_{i,p}}\right)
\left(\prod_{l\in  \mathcal{T}_{v,\{i_d\},\{ {\mathcal{C} }(l) \} }  } \delta_{ \cD   }^{l,{\mathcal{C}}(l)}\right) \crcr
& \quad \times \left( \prod_{d=1}^k \prod_{c=1}^D \delta_{n^c_{i_d,q+1}\, m_{i_{\tau_c(d)},q}^c}\right) \; , \nonumber
\end{align}
where $\tau_{\cD}=(\tau_1, \dots \tau_D)$  runs over $D$-uples of permutations of $k$ elements 
and $ \mathrm{Wg}(\pi,N)$ is the Weingarten function \cite{collins,ColSni}.
\end{theorem}

This theorem will be proved in the section \ref{secLVE}. We see from eq. \eqref{eq:cumulantfin} that the cumulants are 
linear combinations of {\it trace invariant operators} \cite{universality} 
(pairwise identifications of the indices of $T$ and $\bar T$)
of the type:
\[
 \prod_{d=1}^k \prod_{c=1}^D\delta_{n_{ d }^c \bar{n}_{\rho_c ( d)}^c} \; .
\]
Such an operator is specified by $D$ permutations $(\rho_1,\dots \rho_D)$. It is canonically represented as an edge colored graph
\cite{RGurau}. The graph is obtained as follows: we draw a black and a white vertex for every $d$ and, for all $d$ and $c$,
we connect the black vertex $d$ to the white vertex $\rho_c(d)$ by an edge of color $c$.
The $\pi$ and $\bar \pi$ permutations in eq. \eqref{eq:cumulantfin} are just trivial relabellings of the vertices.

\begin{figure}[ht]
\centering
\includegraphics[height=5.5cm]{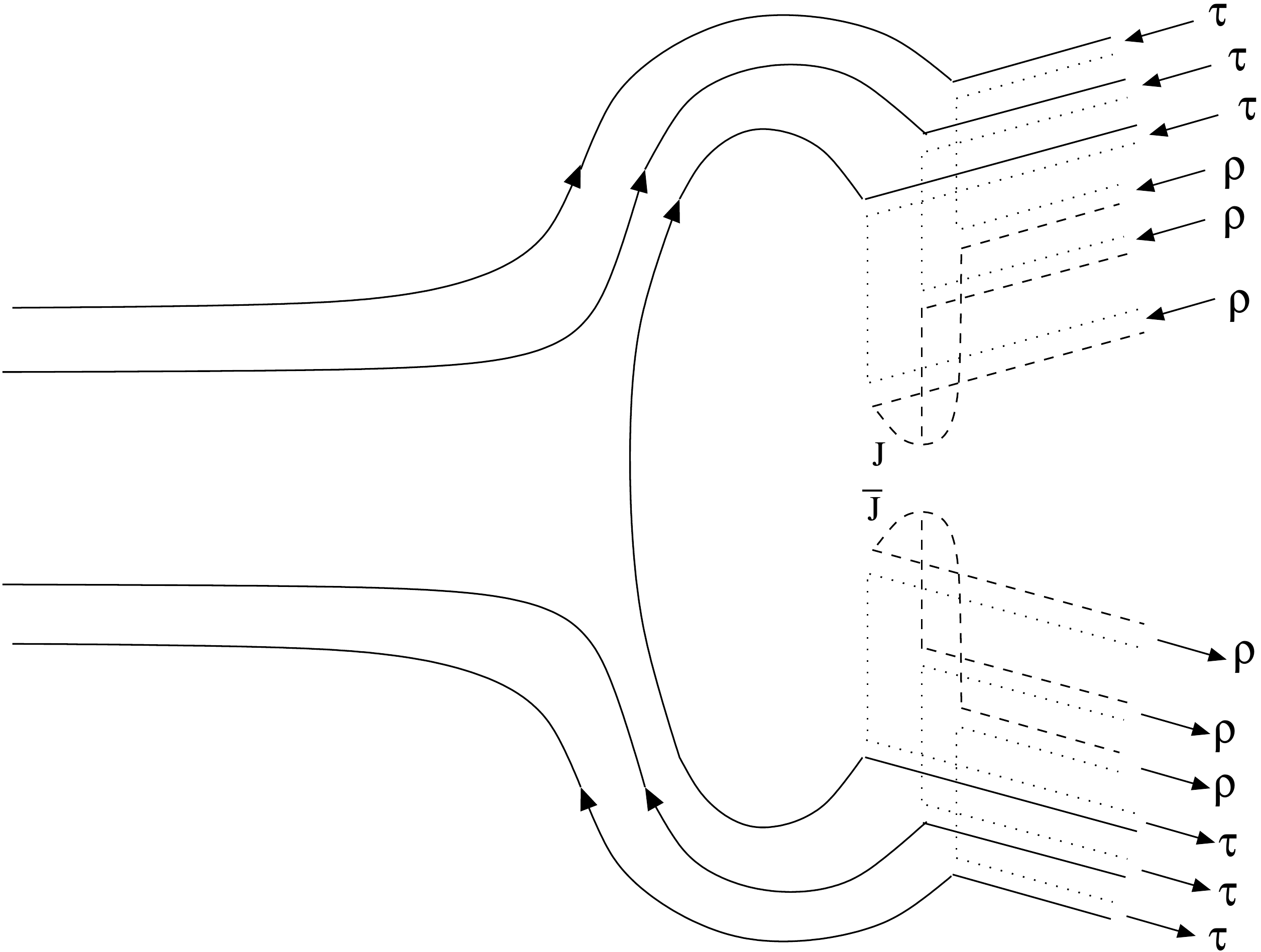}
\caption{Detailed representation of the vertex with external strands.}
\label{fig:fullvertex}
\end{figure}

The $\prod_{d=1}^k \delta_{n_d^c \, m_{\rho_c(d)}^c}\ \delta_{n^c_{i_d,q+1}\, m_{i_{\tau_c(d)},q}^c}$ factors in eq. \eqref{eq:cumulantfin} and eq. \eqref{cumulants0}
represent new contractions between the indices of the resolvents $R(\sigma)$ and the $J$ and $\bar J$ tensors. We call these contractions
\emph{external strands}. The permutations
$\tau$ encode contractions exclusively between the indices of resolvents, while the permutations $\rho$ encode contractions exclusively
between the indices of $J$ and $\bar J$. We represent each such contraction as a ribbon edge, like in Figure \ref{fig:fullvertex}. For convenience we inserted a third category of 
strands (represented as dotted in Figure \ref{fig:fullvertex}) in between the 
solid and the dashed strands. 

\begin{figure}[ht]
\centering
\includegraphics[height=8.5cm]{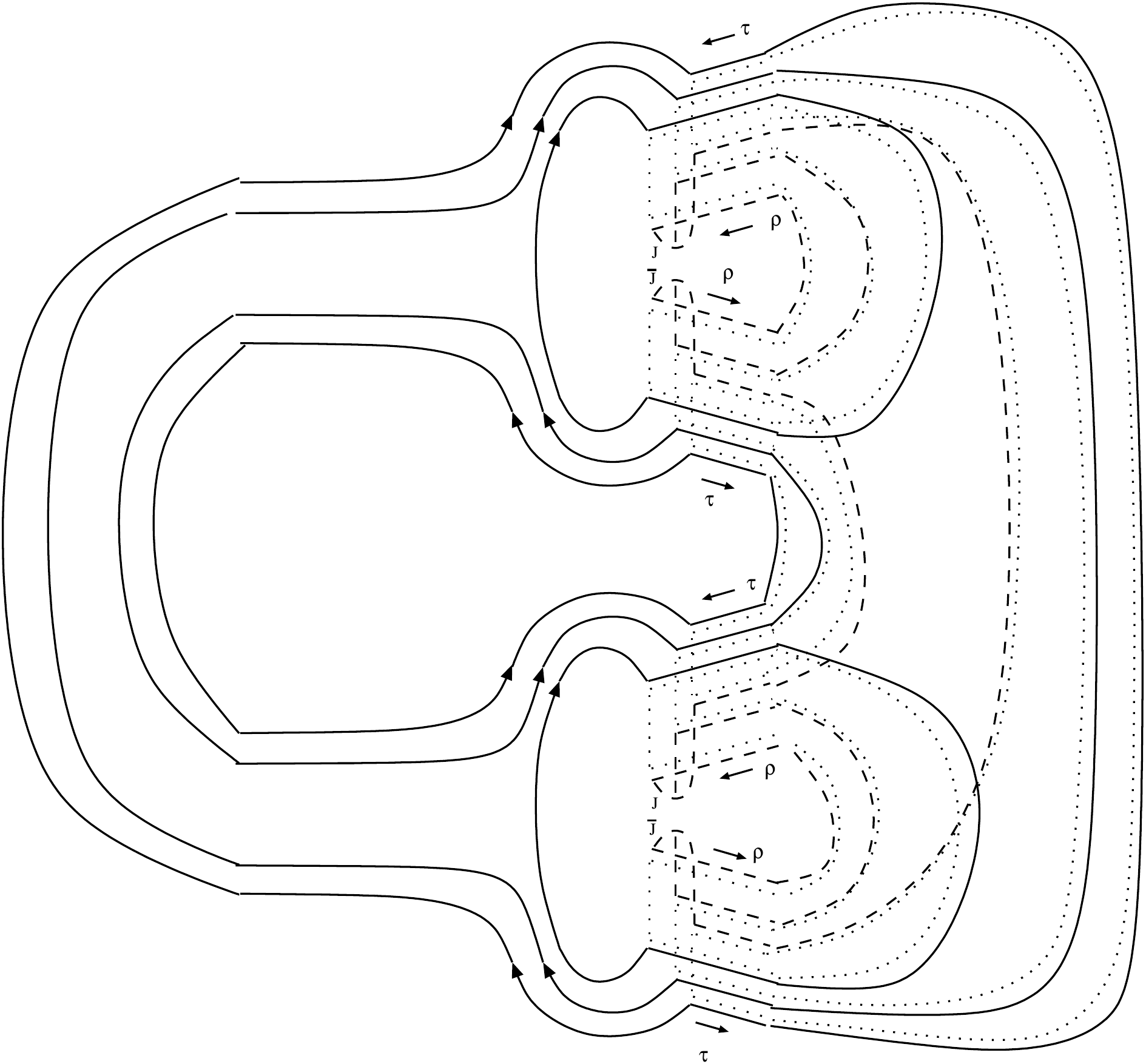}
\caption{Tree with external strands.}
\label{graph22}
\end{figure}

The permutations $\tau_c$ are represented as ribbon edges having a solid strand and a dotted strand, while the $\rho_c$ permutations
have a dashed and a dotted strand. The cycles of the permutation $\tau_c\rho_c^{-1}$ are the closed circuits made of dotted 
strands. A typical example of a tree with external strands is presented in Figure \ref{graph22}.

The graph of each trace invariant in $J$ and $\bar J$ in the expansion \eqref{eq:cumulantfin} is immediately read off our graphical representation: it is the graph 
associated to the permutation $\rho_{\cD}$, hence it is the graph made by the dashed strands.

If one splits the interactions such that $1\notin \cC$, which (as already mentioned) is always possible, 
the color $1$ factors completely. The cumulants rewrite then in terms of:
\begin{itemize}
 \item \emph{reduced resolvents}:
\begin{align}\label{eq:reducedres}
 \res(\sigma)=\left[\mathbf{1}^{\cD\setminus \{1\} }+\sqrt{\frac{\lambda}{ N^{D-1}}} 
 \left(  \sum_{\mathcal{C}}\mathbf1^{ \cD \setminus \{1\} \setminus \cC  }
  \otimes(\sigma^{\mathcal{C}}-\sigma^{{\mathcal{C}}*}) \right)\right]^{-1} \;,
\end{align}
which are operators on a vector space of dimension $N^{D-1}$ corresponding to the indices $\mathbf{n}\equiv(n^2 \dots n^D)$ of colors 
different form $1$. 
\item \emph{reduced edge contractions}:
\begin{align}\label{eq:reduceddelta}
 & \delta_{ \cD \setminus \{1\} }^{l_{ij},{\mathcal{C}}(l_{ij})}=  \\
 &\; =\left( \delta_{m^{\cD\setminus \{1\} \setminus \cC }_{i,q} n^{\cD\setminus \{1\} \setminus \cC }_{i,q+1} } \right) 
    \delta_{ m^{\cC}_{i,q} n^{\cC}_{j,p+1} }      \delta_{ m^{\cC}_{j,p} n^{\cC}_{i,q+1} }
 \left( \delta_{m^{\cD\setminus \{1\} \setminus \cC }_{j,p} n^{\cD\setminus \{1\} \setminus \cC }_{j,p+1} } \right) \; . \nonumber
\end{align}
\end{itemize} 

\begin{corollary}\label{th:LVE1} The cumulants of the measure $\mu$ are alternatively given by:
\begin{align}\label{eq:cumulantfin1}
&\kappa(T_{n_1}\bar{T}_{\bar{n}_1}...T_{n_k}\bar{T}_{\bar{n}_k}) = \crcr 
& \quad =\sum_{\pi, \bar\pi}\sum_{ \rho_{\cD\setminus \{1\} } }\mathfrak{K}(  \rho_{\cD\setminus \{1\} } )\prod_{d=1}^k \delta_{n_{\pi(d)}^1 
\bar{n}_{\bar\pi(d)}^1}\prod_{c=2}^D\delta_{n_{\pi(d)}^c \bar{n}_{\rho_c\bar\pi(d)}^c} \;,
\end{align}
where $ \rho_{\cD\setminus \{1\} }=(\rho_2...\rho_D)$ runs over collections of $D-1$ permutations over $k$ elements, 
$\pi$ and $\bar\pi$ are permutations over $k$ elements, 
and $\mathfrak{K}( \rho_{\cD\setminus \{1\} } )$ is:  
\begin{align}\label{cumulants01}
 \mathfrak{K}( \rho_{\cD\setminus \{1\} })= &
\sum_{v\geq k}\frac1{v!}\ \frac{(-2\lambda)^{v-1}}{N^{(D-1)(k+v-1)+k-v}} \sum_{\tau_{\cD\setminus \{1\} } } 
\left( \prod_{c\neq1} \mathrm{Wg}(N,\tau_c \rho_c^{-1})\right)\crcr
&\quad \times \sum_{\mathcal{T}_{v,\{i_d\},\{{\mathcal{C}}(l)\}}}
\int_0^1 \left(\prod_{l_{ij}\in T_v}du_{ij}\right)\int d\mu_{T_v,u}(\sigma)  \sum_{n,m}\\
&\quad \times\left(\prod_{i=1}^v\prod_{p=1}^{\mathrm{res}(i)}  \res(\sigma^i)_{\mb n_{i,p}\mb m_{i,p}}\right)
\left(\prod_{l\in  \mathcal{T}_{v,\{i_d\},\{ {\mathcal{C} }(l) \} }  } \delta_{ \cD \setminus \{1\} }^{l,{\mathcal{C}}(l)}\right) \crcr
& \quad \times \left( \prod_{d=1}^k \prod_{c=2}^D \delta_{n^c_{i_d,q+1}\, m_{i_{\tau_c(d)},q}^c}\right) \; , \nonumber
\end{align}
where $\tau_{\cD\setminus \{1\} } = (\tau_2,\dots \tau_{D}) $ runs over $D$-uples of permutations of $k$ elements.
\end{corollary}

Eq. \eqref{cumulants01} is not just a trivial evaluation of eq. \eqref{cumulants0}: both the scaling with $N$
and the number of sums in the first lines of the two equations differ. In fact this equation represents a very different repackaging 
of the terms.

The advantage of this second representation resides in the fact that the expansion in eq. \eqref{eq:cumulantfin1}
is written in terms of trace invariants which are trivial on the color $1$: the permutation associated to the indices of color $1$ 
is the identity permutation  $e(d)=d$. If we represent such invariants as edge color graphs, 
the edges of color $1$ always connect the white and the black vertex with the same $d$.

\medskip

\noindent{\bf Mixed Expansion.} The LVE expansion of the cumulants can be refined to an expansion in \emph{trees decorated by loop edges}. 
The loop edges are of the same nature as the tree edges: they have colors $\cC$, they connect vertices and they are adjacent to corners. 
Like tree edges, loop edges represent identifications  of indices of the adjacent resolvents. We represent them as edges with solid strands.

A tree $\cT$ decorated by loop edges $\cL$ (and external strands $\tau_{\cD}$) is a graph. All its edges have strands, and all its vertices are fat. 
We call the closed circuits of solid and dotted strands \emph{faces}. The faces have a color $c\in \cD$.
There are three categories of faces:
\begin{itemize}
 \item faces made of solid strands which do not reach any mark. We call them \emph{internal} faces and denote their number $F_{\rm int}(\cT,\cL)$.
 \item faces made of solid strands which reach at least a mark. When reaching a mark the face follows the permutation $\tau$.
       We call such faces \emph{external} and denote their number $F_{\rm ext}(\cT,\cL,\tau_{\cD})$.
 \item faces made of the dotted strands. We call them $\tau\rho^{-1}$-faces as they track the permutation $\tau\rho^{-1}$. 
       The number of $\tau\rho^{-1}$-faces is the number of cycles of the permutation $\tau \rho^{-1}$, which we denote $C(\tau\rho^{-1})$. They appear explicitly
       in the Weingarten function.
\end{itemize}

The \emph{mixed expansion} of the cumulants is 
\begin{theorem}\label{th:mixed}
The cumulants of $\mu$ write:
\begin{align}\label{eq:mixedexp}
&\mathfrak{K}(\rho_{\cD})=
\sum_{v\geq k}\frac1{v!}\ \frac{(-2\lambda)^{v-1}}{N^{(D-1)(k+v-1) }} \sum_{\tau_{\cD}} \left( \prod_{c = 1}^D 
\mathrm{Wg}(N,\tau_c \rho_c^{-1})\right)   \sum_{\mathcal{T}_{v,\{i_d\},\{{\mathcal{C}}(l)\}}} 
   \crcr
&\times
  \Bigg{[}  \sum_{q=0}^L \left(\frac{-2\lambda}{N^{D-1}}\right)^q  \frac{1}{q!} \sum_{\cL, |\cL| = q}
  N^{F_{\rm int} (\cT,\cL )+F_{ \rm ext } (\cT,\cL,\tau_{ \cD} ) } \crcr
&  \qquad \qquad\qquad\qquad\times \int_0^1 \left(\prod_{l_{ij}\in T_v}du_{ij}\right)   \prod_{l\in \cL } w_{i(l)j(l)}
  \crcr
& \qquad + \left(\frac{-2\lambda}{N^{D-1}}\right)^{L+1}  \frac{1}{L!}  \sum_{ \cL, |\cL| = L+1 } \crcr
& \qquad \qquad\qquad\qquad\times   \int_0^1 dt  \; (1-t)^L  \int_0^1 \left(\prod_{l_{ij}\in T_v}du_{ij}\right)  
   \left( \prod_{l\in \cL }   w_{i(l)j(l)} \right) \crcr
& \qquad \times
\int d\mu_{T_v,u}(\sigma) \sum_{m,n} \left(\prod_{i=1}^v\prod_{p=1}^{\mathrm{res}(i)}  R( \sqrt{t}\sigma^i)_{n_{i,p} m_{i,p}}  \right)
\crcr
&\qquad \times \left(\prod_{l\in \cL }\delta_{\cD }^{l,{\mathcal{C}}(l)} \right)   \left(\prod_{l\in T_v}\delta_{\cD}^{l,{\mathcal{C}}(l)}\right)
\left( \prod_{d=1}^k \prod_{c=2}^D \delta_{n^c_{i_d,q+1}\, m_{i_{\tau_c(d)},q}^c}\right) \Bigg{]} 
\; .
\end{align}
\end{theorem}

This theorem is proved in section \ref{sec:mixed}.

\medskip

\noindent{\bf Absolute convergence and bounds.} From now on we only study the case $|\cC|\le  D/2$, namely we always separate an interaction with the intermediate
matrix field of minimal size. This is of course always possible. We will study the analyticity properties and the scaling with $N$ of the cumulants starting 
from the mixed expansion in eq. \eqref{eq:mixedexp}.

\begin{theorem}\label{th:conv}
 The series in (\ref{eq:mixedexp}) is absolutely convergent for $\lambda\in[0,\frac1{8\mathcal N_{\mathcal{Q}}})$.
 In this domain the cumulants obey the bound 
 \[
 | \mathfrak{K}(\rho_{\cD}) | \le N^{D - 2(D-1) k -  \mathfrak{C}(\rho_{\cD})   }  K(\lambda) \; ,
 \]
 for some $K$ depending only on $\lambda$, and the rescaled second cumulant $N^{D-1}\mathfrak{K}({1}_{\mathcal D})$ admits a finite limit at large $N$. Hence the measure $\mu$ is properly uniformly bounded.
\end{theorem}
This theorem will be proved in section \ref{absconv}. This shows that the measure $\mu$ obeys the universality theorem.

\medskip

\noindent{\bf Uniform Borel summability.} We subsequently establish the uniform Borel summability of the cumulants at the origin.
\begin{theorem}\label{th:Borel}
The cumulants can be analytically continued for complex $\lambda=r e^{i\phi}$ with 
$r<\frac1{8\mathcal N_{\mathcal{Q}}}\left(\mathrm{cos}\frac\phi2\right)^2$. In this domain they obey the bound:
\begin{align}\label{eq:Borelbound}
 | \mathfrak{K}(\rho_{\cD}) | \le N^{D - 2(D-1) k -  \mathfrak{C}(\rho_{\cD})   } 
 K\left( \frac{|\lambda|}  { \left(\cos \phi /2 \right)^2 } \right) \;,
\end{align}
and are Borel summable in $\lambda$ uniformly in $N$.
\end{theorem}
This theorem will be proved in section \ref{BOREL}.

\medskip

\noindent{\bf The $1/N$ expansion.} 

Furthermore, the mixed expansion in eq. \eqref{eq:mixedexp} is the non perturbative $1/N$ expansion of the cumulants in the following sense
\begin{corollary}\label{th:1/N}
 The rest term in the mixed expansion is analytic 
 in the domain $r<\frac1{8\mathcal N_{\mathcal{Q}}}\left(\mathrm{cos}\frac\phi2\right)^2$ and in this domain it admits the bound:
 \begin{align*}
  & 
\sum_{v\geq k}\frac1{v!}\ \frac{(-2\lambda)^{v-1}}{N^{(D-1)(k+v-1) }} \sum_{\tau_{\cD}} \left( \prod_{c = 1}^D 
\mathrm{Wg}(N,\tau_c \rho_c^{-1})\right)   \sum_{\mathcal{T}_{v,\{i_d\},\{{\mathcal{C}}(l)\}}} 
   \crcr
&\qquad \times
   \left(\frac{-2\lambda}{N^{D-1}}\right)^{L+1}  \frac{1}{L!}  \sum_{ \cL, |\cL| = L+1 }  \int_0^1 dt  \; (1-t)^L  \crcr
& \qquad  \times   \int_0^1 \left(\prod_{l_{ij}\in T_v}du_{ij}\right)  
   \left( \prod_{l\in \cL }   w_{i(l)j(l)} \right) \crcr
& \qquad \times
\int d\mu_{T_v,u}(\sigma) \sum_{m,n} \left(\prod_{i=1}^v\prod_{p=1}^{\mathrm{res}(i)}  R( \sqrt{t}\sigma^i)_{n_{i,p} m_{i,p}}  \right)
\crcr
&\qquad \times \left(\prod_{l\in \cL }\delta_{\cD }^{l,{\mathcal{C}}(l)} \right)   \left(\prod_{l\in T_v}\delta_{\cD}^{l,{\mathcal{C}}(l)}\right)
\left( \prod_{d=1}^k \prod_{c=2}^D \delta_{n^c_{i_d,q+1}\, m_{i_{\tau_c(d)},q}^c}\right) \Bigg{]} \crcr
& \le N^{D - (D-1)k - (L+1) \left(\frac{D}{2}-1\right) }
\left( \frac{|\lambda|}  { \left(\cos \phi /2 \right)^2 } \right)^{L+k}
K'\left( \frac{|\lambda|}  { \left(\cos \phi /2 \right)^2 } \right) 
\; .
 \end{align*}
for some bounded function $K'$.
\end{corollary}

\section{Loop Vertex Expansion}\label{secLVE}

In this section we prove Lemma \ref{lem:lemmaLVE}, Theorem \ref{th:LVE} and Corollary \ref{th:LVE1}.

\subsection{Intermediate field representation}\label{subsecIFR}

The Hubbard Stratonovich intermediate field representation relies on the observation that, 
for any complex numbers $Z_1,Z_2$, 
\[
  \int \frac{d\bar z dz}{2 \imath \pi} \;  e^{-z\bar z - z Z_1 + \bar z Z_2} = e^{-Z_1Z_2} \; .
\]
We will now apply this formula for the quartic interaction terms. We have:
\begin{align*}
& e^{-N^{D-1}\lambda \tr_{\cC} \left[  [ {\bf T}^{\vee} \cdot_{\cD\setminus \cC} {\bf T} ] \cdot_{\cC}
  [ {\bf T}^{\vee} \cdot_{\cD\setminus \cC} {\bf T} ] \right] } = \crcr
& \; \;  = \prod_{ n^{\cC} \bar n^{\cC} m^{\cC} \bar m^{\cC} }
    e^{-N^{D-1}\lambda    [ {\bf T}^{\vee} \cdot_{\cD\setminus \cC} {\bf T} ]_{ \bar n^{\cC}   n^{\cC}   }
         \delta_{n^{\cC} \bar m^{\cC}} \delta_{ m^{\cC} \bar n^{\cC}}
  [ {\bf T}^{\vee} \cdot_{\cD\setminus \cC} {\bf T} ]_{ \bar m^{\cC}  m^{\cC} } } 
  \crcr
& 
  \; \;  = \prod_{ n^{\cC} \bar n^{\cC}  }
    e^{-N^{D-1}\lambda    [ {\bf T}^{\vee} \cdot_{\cD\setminus \cC} {\bf T} ]_{ \bar n^{\cC}   n^{\cC}   }
  [ {\bf T}^{\vee} \cdot_{\cD\setminus \cC} {\bf T} ]_{ n^{\cC}  \bar n^{\cC} } } 
\crcr
  &=  \prod_{ n^{\cC} \bar n^{\cC} }   \Bigg(
   \int \frac{d\sigma_{n^{\cal C} \bar n^{\cal C}}^{\mathcal{C}}d\bar{\sigma}_{n^{\cal C} \bar n^{\cal C}}^{\mathcal{C}}}{2\imath\pi}
    \quad e^{- \sigma_{ \bar n^{\cC}  n^{\cC} }^{\cC}  \bar \sigma_{  \bar n^{\cC}  n^{\cC} }^{\cC}   } 
    \crcr 
 & \qquad\times 
   e^{-\sqrt{\lambda N^{D-1} }   [ {\bf T}^{\vee} \cdot_{\cD\setminus \cC} {\bf T} ]_{ \bar n^{\cC}  n^{\cC}  } \sigma^{\cC}_{ \bar n^{\cC} n^{\cC} }       }
  \;\;
    e^{ \sqrt{\lambda N^{D-1} }   [ {\bf T}^{\vee} \cdot_{\cD\setminus \cC} {\bf T} ]_{ n^{\cC}  \bar n^{\cC} } \bar \sigma^{\cC}_{ \bar n^{\cC} n^{\cC}  }    }  \Bigg)  
  \crcr
   &= \int \left(\prod_{a^{\cC}b^{\cC}} \frac{d\sigma_{a^{\cC}b^{\cC}}^{\mathcal{C}}d\bar{\sigma}_{a^{\cC}b^{\cC}}^{\mathcal{C}}}{2 \imath \pi}\right) 
   e^{- \tr_{\cC} \left[ \sigma^{\cC*} \sigma^{\cC} \right] } \crcr
   & \qquad \qquad  \times e^{  - \sqrt{\lambda N^{D-1}} 
    \sum_{n \bar n} \bar T_{\bar n} \left[  \mathbf1^{\cD\setminus \cC} 
   \otimes(\sigma^{\mathcal{C}}-\sigma^{{\mathcal{C}}*}) \right]_{\bar n n}  T_{ n } } \; .
\end{align*}
The generating function is then:
\begin{align*}
& Z(J, \bar{J}) = \crcr
& =\int  \left(\prod_{n} N^{D-1}\frac{dT_{n} d \bar{T}_{n } } {2\imath \pi}\right) 
\left(\prod_{\mathcal{C}}\prod_{a^{\cC}b^{\cC}}  \frac{d\sigma_{  a^{\cC}b^{\cC}  }^{\mathcal{C}}d\bar{\sigma}_{  a^{\cC}b^{\cC}  }^{\mathcal{C}}}{2\pi}\right)
e^{-\sum_{\mathcal{C}}\tr\sigma^{\mathcal{C}}\sigma^{{\mathcal{C}}*}} \crcr
&\;\; \times e^{  -N^{D-1}\sum_{n \bar n} \bar T_{\bar n }    \left[ {\mathbf1}^{\cD} \ +\ \sqrt{\frac{\lambda}{ N^{D-1}}} 
\left(  \sum_{\mathcal{C}}\mathbf1^{ \cD \setminus \cC }\otimes(\sigma^{\mathcal{C}}-\sigma^{{\mathcal{C}}*}) \right)\right]_{\bar n; n} 
   T_{n} } \crcr
& \;\; \times  e^{ \sum_{n} T_{n} \bar{J}_{n} + \sum_{\bar n}\bar{T}_{\bar{n} }J_{\bar{n} }}  \; .
\end{align*}
The integral over $T$ and $\bar T$ is now Gaussian, and a direct computation leads to eq. \eqref{eq:ZLVE}.

\subsection{Forest formula}\label{subsecFF}

To simplify notations we sometimes drop the superscript $\cC$ on the (multi) indices
of $\sigma^{\cC}$. According to our equation \eqref{eq:forest}, the logarithm of $Z(J,\bar J)$ is:
\begin{align*}
\lnz (J,\bar J)&=\sum_{v\geq 1}\frac1{v!} \sum_{T_v}\int_0^1 \left(\prod_{l_{ij}\in T_v}du_{ij}\right)\int d\mu_{T_v,u}(\sigma)\crcr
&\times\left[\prod_{l_{ij}\in T_v}\left(
\sum_{\cC,ab }\left(
\frac{\dr}{\dr\sigma_{ab}^{i\ {\mathcal{C}}}}\frac{\dr}{\dr\bar\sigma_{ab}^{j\ {\mathcal{C}}}} 
+ \frac{\dr}{\dr\sigma_{ab}^{j\ {\mathcal{C}}}}\frac{\dr}{\dr\bar\sigma_{ab}^{i\ {\mathcal{C}}}}
\right)\right)
\right]\crcr
&\times\prod_{i=1}^v \left(-\tr\ \mathrm{ln}\left(\mathbf1^{\cD} + A(\sigma^i)\right)+N^{1-D}\bar J_n\left(\mathbf1^{\cD} + A(\sigma^i)\right)^{-1}_{nm}J_m\right),
\end{align*}
where $T_v$ are combinatorial trees with $v$ vertices and the interpolated Gaussian measure $d\mu_{T_v, u}$ is degenerated over $\cC$:
\begin{align*}
 \int F(\sigma) \; d\mu_{T_v, u} = \left[e^{\sum_{i,j}w_{ij}\sum_{ \cC,ab }\left(
\frac{\dr}{\dr\sigma_{ab}^{i\ {\mathcal{C}} }}\frac{\dr}{\dr\bar\sigma_{ab}^{j\ {\mathcal{C}}}} 
\right)}F(\sigma)\right]_{\sigma=0} \; ,
\end{align*}
and $w_{ij}$ is defined in equation \eqref{eq:wdef}. Expanding the product over $i$ we get:
\begin{align}\label{eq:intermediar}
&\lnz (J,\bar J)=\sum_{v\geq 1}\frac1{v!} \sum_{T_v}\int_0^1 \left(\prod_{l_{ij}\in T_v}du_{ij}\right)\int d\mu_{T_v,u}(\sigma)\crcr
& \;\; \times\left[\prod_{l_{ij}\in T_v}\left(
\sum_{ \cC,ab }\left(
\frac{\dr}{\dr\sigma_{ab}^{i\ {\mathcal{C}}}}\frac{\dr}{\dr\bar\sigma_{ab}^{j\ {\mathcal{C}}}} 
+ \frac{\dr}{\dr\sigma_{ab}^{j\ {\mathcal{C}}}}\frac{\dr}{\dr\bar\sigma_{ab}^{i\ {\mathcal{C}}}}
\right)\right)
\right]\sum_{k=1}^v\ \sum_{i_1<...< i_k} \crcr
& \;\; \times \prod_{d=1}^k\ N^{1-D}\bar J_n\left(\mathbf1^{\cD} + A(\sigma^{i_d})\right)^{-1}_{nm}J_m
\prod_{i\neq i_1..i_k}-\tr\ \mathrm{ln}\left(\mathbf1^{\cD} + A(\sigma^i)\right) \;.
\end{align}
The logarithm of $Z(J,\bar J)$ is then a sum over trees $T_{v,k,\{i_d\}}$ with $k$ marked vertices (the $J\bar J$ vertices) 
and $v-k$ regular vertices. The sum over ${\mathcal{C}}$ gives us a sum over trees with colored edges, each coloring corresponding 
to a set ${\mathcal{C}}\in\mathcal{Q}$.  

Before taking into account the action of the derivatives, to each marked vertex $i_d$ of the tree $T_{v,k,\{i_d\}}$ is associated a resolvent 
operator $R(\sigma^{i_d})=\left(\mathbf1^{\cD} + A(\sigma^{i_d}) \right)^{-1}$ (and a pair $J, \bar J$), and 
to each unmarked vertex $i$ is associated a $ -\tr\ \mathrm{ln}\left(\mathbf1 + A(\sigma^i)\right)  $ factor.

We now have to evaluate the action of the derivatives: 
\begin{align}\label{eq:derivative}
 & \frac{\partial}{\partial \sigma^{i \ \cC}_{a^{\cC} b^{\cC} }} \left[ R(\sigma^{i}) \right]_{nm} = \crcr
 & = 
  \frac{\partial}{\partial \sigma^{i \ \cC}_{a^{\cC} b^{\cC} }} \left[ \sum_{q=0}^{\infty} \left( - \sqrt{\frac{\lambda}{N^{D-1}}} \right)^q
   \left(    \sum_{\mathcal{C}}\mathbf1^{ \cD \setminus \cC }\otimes(\sigma^{i \ \cC }-\sigma^{ i \ \cC *})   \right)^q \right]_{nm}  \crcr
 & = \left( - \sqrt{\frac{\lambda}{N^{D-1}}} \right) \sum_{a^{\cD\setminus \cC} b^{\cD\setminus \cC} }\sum_{q_1,q_2=0}^{\infty} \crcr
& \; \times \left( - \sqrt{\frac{\lambda}{N^{D-1}}} \right)^{q_1}
   \left(    \sum_{\mathcal{C}}\mathbf1^{ \cD \setminus \cC }\otimes(\sigma^{i \ \cC }-\sigma^{ i \ \cC *})   \right)^{q_1}_{n a} \crcr
& \; \times \left( - \sqrt{\frac{\lambda}{N^{D-1}}} \right)^{q_2}
   \left(    \sum_{\mathcal{C}}\mathbf1^{ \cD \setminus \cC }\otimes(\sigma^{i \ \cC }-\sigma^{ i \ \cC *})   \right)^{q_2}_{b m}  
     \delta_{a^{\cD\setminus \cC} b^{\cD \setminus \cC} } \crcr
& = \left( - \sqrt{\frac{\lambda}{N^{D-1}}} \right) \sum_{ a^{\cD\setminus \cC} b^{\cD\setminus \cC}  } 
   \left[ R(\sigma^{i}) \right]_{n a } \delta_{a^{\cD\setminus \cC} b^{\cD \setminus \cC} }  \left[ R(\sigma^{i}) \right]_{b m} \crcr
 &  \frac{\partial}{\partial \sigma^{i \ \cC}_{a^{\cC} b^{\cC} }} \left( - \tr\ln[  \left(\mathbf1 + A(\sigma^i)\right)  ] \right) = \crcr
 & =  \frac{\partial}{\partial \sigma^{i \ \cC}_{a^{\cC} b^{\cC} }} \left[  \sum_{q=1}^{\infty}\frac{(-1)^q}{q} 
    \left(   \sqrt{\frac{\lambda}{N^{D-1}}} \right)^q \tr
   \left(    \sum_{\mathcal{C}}\mathbf1^{ \cD \setminus \cC }\otimes(\sigma^{i \ \cC }-\sigma^{ i \ \cC *})   \right)^q
 \right] \crcr
& =  \left( - \sqrt{\frac{\lambda}{N^{D-1}}} \right) \sum_{q=0}^{\infty} \left( -  \sqrt{\frac{\lambda}{N^{D-1}}} \right)^q \crcr
& \;\; \times \sum_{a^{\cD\setminus \cD} b^{\cD \setminus \cC}}  \delta_{a^{\cD \setminus \cD} b^{\cD \setminus \cC} }
 \left[ \left(    \sum_{\mathcal{C}}\mathbf1^{ \cD \setminus \cC }\otimes(\sigma^{i \ \cC }-\sigma^{ i \ \cC *})   \right)^q \right]_{ba} \crcr
& = \left( - \sqrt{\frac{\lambda}{N^{D-1}}} \right) \sum_{a^{\cD\setminus \cD} b^{\cD \setminus \cC}}  \delta_{a^{\cD \setminus \cD} b^{\cD \setminus \cC} }
     \left[ R(\sigma^{i}) \right]_{b a }   \; ,
\end{align}
and similarly for $\bar \sigma$. A couple of derivative operators (i.e. an edge) adds a resolvent on each vertex it acts on.
On each vertex, marked or not, acts at least one derivative operator, thus we obtain at least a resolvent per vertex. 
Each vertex is then a partial trace of this resolvents (and a pair $J,\bar J$ if it is marked).

The action of the derivatives acting on a vertex $(p+1)$-th edge hooked to a vertex is the sum of the $p$ positions one can add a resolvent into 
the partial trace of eq. \eqref{eq:derivative}, and induces a well defined ordering of the resolvents at a vertex. 
The sum in \eqref{eq:intermediar} becomes thus a sum over plane trees, with well defined orderings
of the half edges at every vertex, with resolvents associated to the corners.

We have thus expressed $\lnz$ as a sum over plane trees with colored edges and marked vertices $\mathcal{T}_{v,\{i_d\},\{{\mathcal{C}}(l)\}}$. 
The contribution of a tree is a product of resolvents and $J\bar J$ with indices contracted in a certain pattern. 
To the edge $l_{ij}$ connecting the vertices $i$ and $j$, incident at the corners $q$ and $q+1$ of the vertex $i$ and $p$ and $p+1$
of the vertex $j$, corresponds the contraction:
\begin{align*}
 \delta_{\cD}^{l_{ij},{\mathcal{C}}(l_{ij})}=  
 \left( \delta_{m^{\cD \setminus \cC }_{i,q} n^{\cD  \setminus \cC }_{i,q+1} } \right) 
    \delta_{ m^{\cC}_{i,q} n^{\cC}_{j,p+1} }      \delta_{ m^{\cC}_{j,p} n^{\cC}_{i,q+1} }
 \left( \delta_{m^{\cD \setminus \cC }_{j,p} n^{\cD \setminus \cC }_{j,p+1} } \right) \; .
\end{align*}
 Collecting everything and taking into account that each edge is the sum of two terms we obtain: 
\begin{align}\label{eq:lnz1}
&\lnz (J,\bar J)=
\sum_{v\geq 1}\frac1{v!}\sum_{k=1}^v\ 
\sum_{\mathcal{T}_{v,\{i_d\},\{{\mathcal{C}}(l)\}}}
\int_0^1 \left(\prod_{l_{ij}\in T_v}du_{ij}\right)\int d\mu_{T_v,u}(\sigma) \sum_{n,m} \crcr
& \qquad \times \left(\prod_{i=1}^v\prod_{p=1}^{\mathrm{res}(i)} R(\sigma^i)_{n_{i,p}m_{i,p}}\right)
\left(\prod_{l\in T_v}\frac{-2\lambda}{N^{D-1}} \ \delta_{\cD}^{l,{\mathcal{C}}(l)}\right) \crcr
& \qquad \times \left(\prod_{d=1}^k\ N^{1-D}\bar J_{n_{i_d,q+1}}J_{m_{i_d,q}}\right) \; ,
\end{align}
where res$(i)$, the number of resolvents associated with the vertex $i$, equals its degree for unmarked vertices, 
and its degree plus one for marked vertices, in the last line $q$ denotes the position of the $J\bar J$ mark on the vertex 
$i_d$, and all the indices $n$ and $m$ are summed.

If we always chose $\cC$ such that $1\notin \cC$, then $A(\sigma)$ is always trivial on the index of color $1$,
\begin{align*}
 A(\sigma) =  1^{\{1\}} \otimes \Bigl( \sqrt{  \frac{\lambda}{N^{D-1} } } \sum_{\cC} {\bf 1}^{\cD\setminus\{1\}\setminus \cC} \otimes
  (\sigma^{\cC}- \sigma^{\cC*})\Bigr) \;,
\end{align*}
and the same holds for $R(\sigma)$, that is $R(\sigma)=\mathbf1^{\{1\}}\otimes\res(\sigma)$ with $\res(\sigma)$ defined in eq.
\eqref{eq:reducedres}. Also, $\delta^{l,\cC(l)}$ writes as 
\[
\delta_{\cD}^{l,\cC(l)}=\delta_{m_{i,q}^1n_{i,q+1}^1}\delta_{m_{j,p}^1n_{j,p+1}^1}\delta_{\cD\setminus\{1\} }^{l,\cC(l)}
\]
with $\delta_{\cD \setminus \{1\} }^{l, \cC(l)} $ defined in eq. \eqref{eq:reduceddelta}. The trace over the index of color $1$ can be explicitly evaluated.
Denoting $\mb n =n^{\cD \setminus \{1\} } = (n^2,\dots n^D)$, we get 
\begin{align} \label{eq:lnz1reduced}
&\lnz (J,\bar J)=
\sum_{v\geq 1}\frac1{v!}\sum_{k=1}^v\ \frac{(-2\lambda)^{v-1}}{N^{(D-1)(k+v-1)+k-v}}  \sum_{\mathcal{T}_{v,\{i_d\},\{{\mathcal{C}}(l)\}}} \int_0^1 \left(\prod_{l_{ij}\in T_v}du_{ij}\right) \\
& \;\times \int d\mu_{T_v,u}(\sigma) \sum_{n,m}   \left(\prod_{l\in T_v}\delta_{\cD\setminus \{1\} }^{l,{\mathcal{C}}(l)}\right)   
\left(\prod_{i=1}^v\prod_{p=1}^{\mathrm{res}(i)}  \res(\sigma^i)_{\mb n_{i,p}\mb m_{i,p}}\right) \crcr
& \; \times \left(\prod_{d=1}^k \bar J_{n_{i_d,q+1}}J_{m_{i_d,q}} \delta_{  n^1_{i_d,q+1}     m^1_{i_d,q}  } \right) 
 \nonumber \;.
\end{align}
The contribution of the strands of color $1$ has been completely factored out. The only trace of the color 
$1$ still subsisting is the contraction of the indices of colors $1$ between the source tensors 
$J$ and $\bar J$ on the same vertex $i_d$ in the last line of the equation above.

\subsection{Cumulants}\label{subsecCUMU}
 
The cumulants are computed by evaluating the derivatives of eq. \eqref{eq:lnz1} with respect to $J$ and $\bar J$. However, in its present form,
eq. \eqref{eq:lnz1} obscures the invariance properties of the cumulants under unitary transformations. 

\begin{figure}[ht]
\centering
\includegraphics[height=5.5cm]{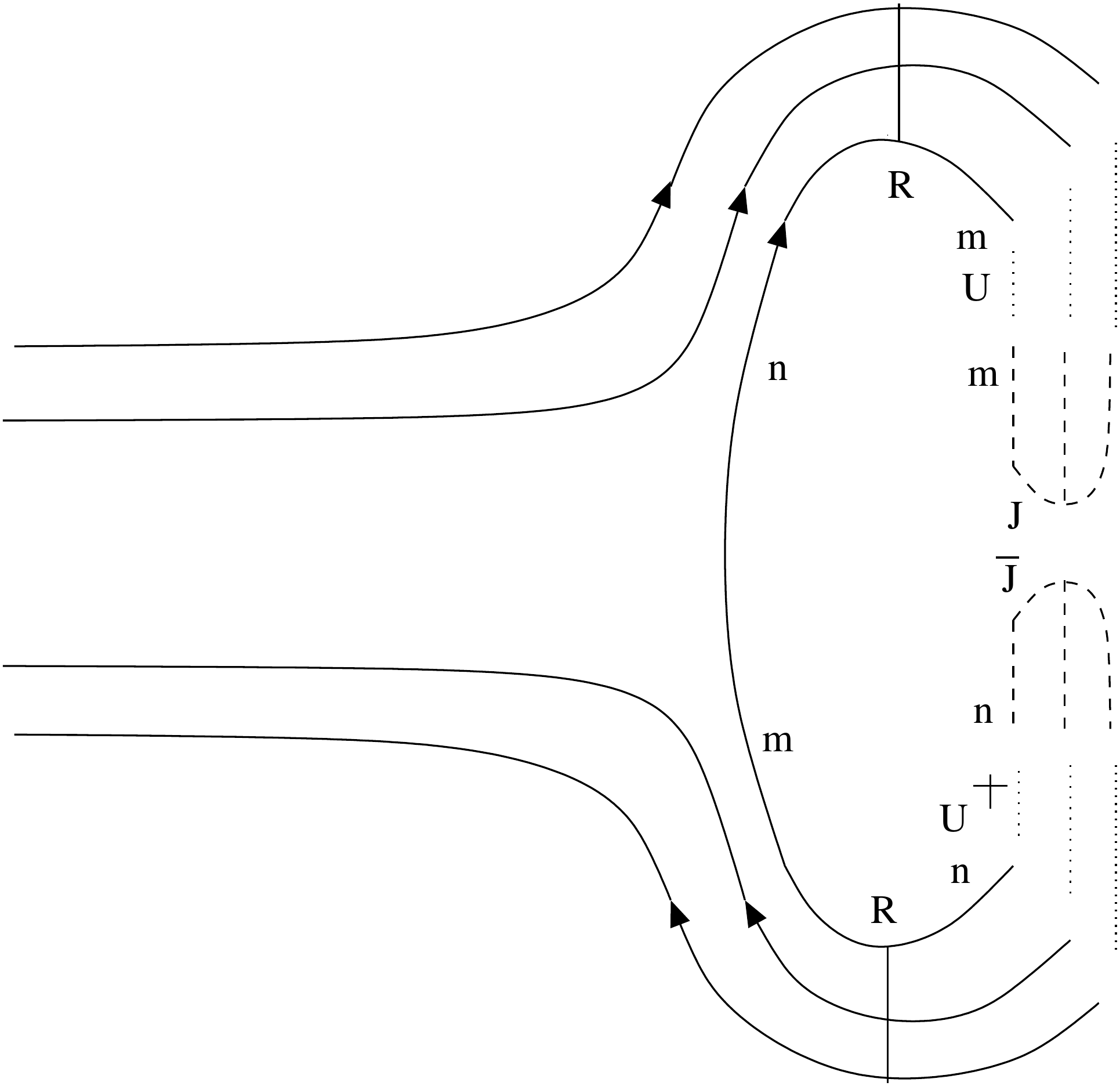}
\caption{The $U$ and $U^{*}$ transformations.}
\label{fig:graph1234}
\end{figure}

To identify the appropriate invariant structure we use a trick introduced in \cite{RGurau}.
Let us consider a set of unitary operators $U^c\in U(N), c\in \cD$. The Gaussian measure is invariant under the change of variables, of unit Jacobian,
\[
\sigma^{i\ \cC }\to(  \otimes_{c\in \cC} U^ {c*}) \ \sigma^{i\ {\mathcal{C}}} \
(  \otimes_{c\in \cC} U^ {c }) \; .
\]
Using $\mathbf1=U^{c*} U^c$, under this change of variables the resolvent changes like 
\begin{align*}
& R(\sigma)\to
\  \left( \bigotimes_{c\in \cD   } U^{c*} \right)  \ R(\sigma)\   \left( \bigotimes_{c\in \cD  } U^{c } \right)  \; .
\end{align*}
 If two resolvent indices are contracted together 
(through an edge contraction $\delta^{l, \cC(l) }_{\cD  }$) an $U^c$ and $U^{c *}$ are multiplied together and drop out.
Thus, the only surviving $U^c$'s are those from indices contracted with the $J, \bar J$ source terms:
\begin{align*}
&\lnz (J,\bar J)= \sum_{v\geq 1}\frac1{v!}\sum_{k=1}^v\ \frac{(-2\lambda)^{v-1}}{N^{(D-1)(k+v-1)}} \crcr
&\qquad \times \sum_{\mathcal{T}_{v,\{i_d\},\{{\mathcal{C}}(l)\}}}
\int_0^1 \left(\prod_{l_{ij}\in T_v}du_{ij}\right)\int d\mu_{T_v,u}(\sigma) \sum_{n,m} \crcr
&\qquad \times \left(\prod_{i=1}^v\prod_{p=1}^{\mathrm{res}(i)} R(\sigma^i)_{ {  n}_{i,p} { m}_{i,p}}\right)
\left(\prod_{l\in T_v}\delta_{\cD }^{l,{\mathcal{C}}(l)}\right) \crcr
& \qquad \times 
\prod_{d=1}^k  \left(  \bar J_{n_d}J_{m_d} 
 \prod_{c=1}^D U^{c*}_{n_d^c n^c_{i_d,q+1}}U^c_{m_{i_d,q}^cm^c_d}\right).\nonumber
\end{align*}
The unitary operators can be added to our graphical representation of Figure \ref{fig:graph123}
by inserting a piece of a dotted strand in between the dashed strands representing the indices of $J$ and the solid strands
representing the indices of the resolvent, as in Figure \ref{fig:graph1234}.

Now, as $\int_{U(N)}dU=1$, we have the trivial equality
\[
\lnz (J,\bar J)=\int_{U(N)}...\int_{U(N)}\lnz (J,\bar J)\ dU^1 \dots dU^D \; ,
\]
and, for each value of $U^c$, we use the previous change of variables. 
The integral over $U(N)$ can be explicitly performed \cite{collins,ColSni}:
\begin{align}\label{eq:uniint}
& \int_{U(N)}dU^c \;  \prod_{d=1}^k U^{c*}_{n_d^c n^c_{i_d,q+1}}  U^c_{m_{i_d,q}^c m^c_d}  \crcr
& \qquad \qquad =\sum_{\rho_c,\tau_c} \mathrm{Wg}(N,\tau_c \rho_c^{-1}) \prod_{d=1}^k \delta_{n_d^c \, m_{\rho_c(d)}^c}\ \delta_{n^c_{i_d,q+1}\, m_{i_{\tau_c(d)},q}^c},
\end{align}
where $\rho_c$ and $\tau_c$ run over all the permutations of $k$ elements. 
The functions $\mathrm{Wg}(N,\pi)$, introduced in \cite{Weingarten}, are known as the Weingarten functions.
They depend only of the cycle structure of the permutation $\pi$ \cite{collins, ColSni}, and if 
$\pi$ has $q$ cycles of lengths $d_1,\dots d_q$, then \cite{collins,ColSni}:
   \begin{align}\label{eq:WeinlargeN}
     &       \mathrm{Wg}(N,(1)) = \frac{1}{N} \; , \qquad     \mathrm{Wg}(N,\pi) = \prod_{j=1}^q V_{d_j} + O(N^{k-2n-2}) \; ,\crcr
     & V_{d} = N^{1-2d} (-1)^{d-1} \frac{1}{d} \binom{2d-2}{d-1} +O(N^{-1-2d}) \; .
     \end{align}
 
The $\prod_{d=1}^k \delta_{n_d^c \, m_{\rho_c(d)}^c}\ \delta_{n^c_{i_d,q+1}\, m_{i_{\tau_c(d)},q}^c}$ factors in eq. \eqref{eq:uniint}
representing new contractions between the indices of the resolvents $R(\sigma)$ and the $J$ and $\bar J$ tensors become the 
\emph{external strands}. The logarithm of $Z$ writes
\begin{align}\label{eq:almosthere}
&\lnz (J,\bar J)=
\sum_{v\geq 1}\frac1{v!}\sum_{k=1}^v\ \frac{(-2\lambda)^{v-1}}{N^{(D-1)(k+v-1) }} 
\sum_{\rho_{\cD},\tau_{\cD}} \left( \prod_{c=1}^D \mathrm{Wg}(N,\tau_c \rho_c^{-1})\right)  \\
& \; \times 
\sum_{n_d,m_d}\left[ \prod_{d=1}^k\left( \bar J_{n_d}J_{m_d} 
\prod_{c=1}^D  \delta_{n_d^c \, m_{\rho_c(d)}^c}\right) \right] \crcr
& \;\times \sum_{\mathcal{T}_{v,\{i_d\},\{{\mathcal{C}}(l)\}}} \int_0^1 \left(\prod_{l_{ij}\in T_v}du_{ij}\right)\int d\mu_{T_v,u}(\sigma) \sum_{n,m}\crcr
&\times\left(\prod_{i=1}^v\prod_{p=1}^{\mathrm{res}(i)} R(\sigma^i)_{  n_{i,p} m_{i,p}}\right)
\left(\prod_{l\in T_v}\delta_{\cD }^{l,{\mathcal{C}}(l)}\right)
\left( \prod_{d=1}^k \prod_{c=1}^D \delta_{n^c_{i_d,q+1}\, m_{i_{\tau_c(d)},q}^c}\right) \nonumber \;.
\end{align}

In this form $\lnz(J,\bar J)$ is explicitly a sum of trace invariants made of $J$ and $\bar J$ tensors, and the 
graph of each such trace invariant is the graph made by the dashed strands.
The coefficient of a trace invariant is a sum over trees (decorated by external strands $\tau_{\cD}$). 
The cumulants are computed by taking the partial derivatives of 
eq. \eqref{eq:almosthere} with respect to the external sources, 
\begin{align*}
\kappa(T_{n_1}\bar{T}_{\bar{n}_1}...T_{n_k}\bar{T}_{\bar{n}_k})
&=\frac{\partial^{2k}}{\partial \bar{J}_{n_1}\partial J_{\bar{n}_1} \dots \partial \bar{J}_{n_k}\partial J_{\bar{n}_k}} \mathrm{ln} Z(J,\bar J)
\Big{\vert}_{J =\bar J =0}\nonumber\\
&=\sum_{\pi, \bar\pi}\sum_{\rho_{\cD}}\mathfrak{K}(\rho_{\cD})\prod_{d=1}^k  
 \prod_{c=1}^D\delta_{n_{\pi(d)}^c \bar{n}_{\rho_c\bar\pi(d)}^c},
\end{align*}
with 
\begin{align}\label{cumulants1}
&\mathfrak{K}(\rho_{\cD})=
\sum_{v\geq k}\frac1{v!}\ \frac{(-2\lambda)^{v-1}}{N^{(D-1)(k+v-1) }} \sum_{\tau_{\cD}} \left( \prod_{c = 1}^D \mathrm{Wg}(N,\tau_c \rho_c^{-1})\right) \crcr
&\times \sum_{\mathcal{T}_{v,\{i_d\},\{{\mathcal{C}}(l)\}}}
\int_0^1 \left(\prod_{l_{ij}\in T_v}du_{ij}\right)\int d\mu_{T_v,u}(\sigma) \sum_{n,m} \\
&\times\left(\prod_{i=1}^v\prod_{p=1}^{\mathrm{res}(i)}  R(\sigma^i)_{ n_{i,p} m_{i,p}}\right)
\left(\prod_{l\in T_v}\delta_{\cD  }^{l,{\mathcal{C}}(l)}\right)
\left( \prod_{d=1}^k \prod_{c=1}^D \delta_{n^c_{i_d,q+1}\, m_{i_{\tau_c(d)},q}^c}\right) \nonumber \; ,
\end{align}
which achieves the proof of Theorem \ref{th:LVE}.

For corollary \ref{th:LVE1}, one follows the same steps with minor adaptations, starting from eq. \eqref{eq:lnz1reduced}. 
Thus the unitary operators $U^c$ act only on the colors $2 \dots D$, the reduced resolvent 
changes like  
\begin{align*}
& \res(\sigma)\to
\  \left( \bigotimes_{c\in \cD \setminus \{1\} } U^{c*} \right)  \ \res(\sigma)\ 
  \left( \bigotimes_{c\in \cD \setminus \{1\} } U^{c } \right)  \; ,
\end{align*}
and so on. 

\begin{figure}[ht]
\centering
\includegraphics[height=5.5cm]{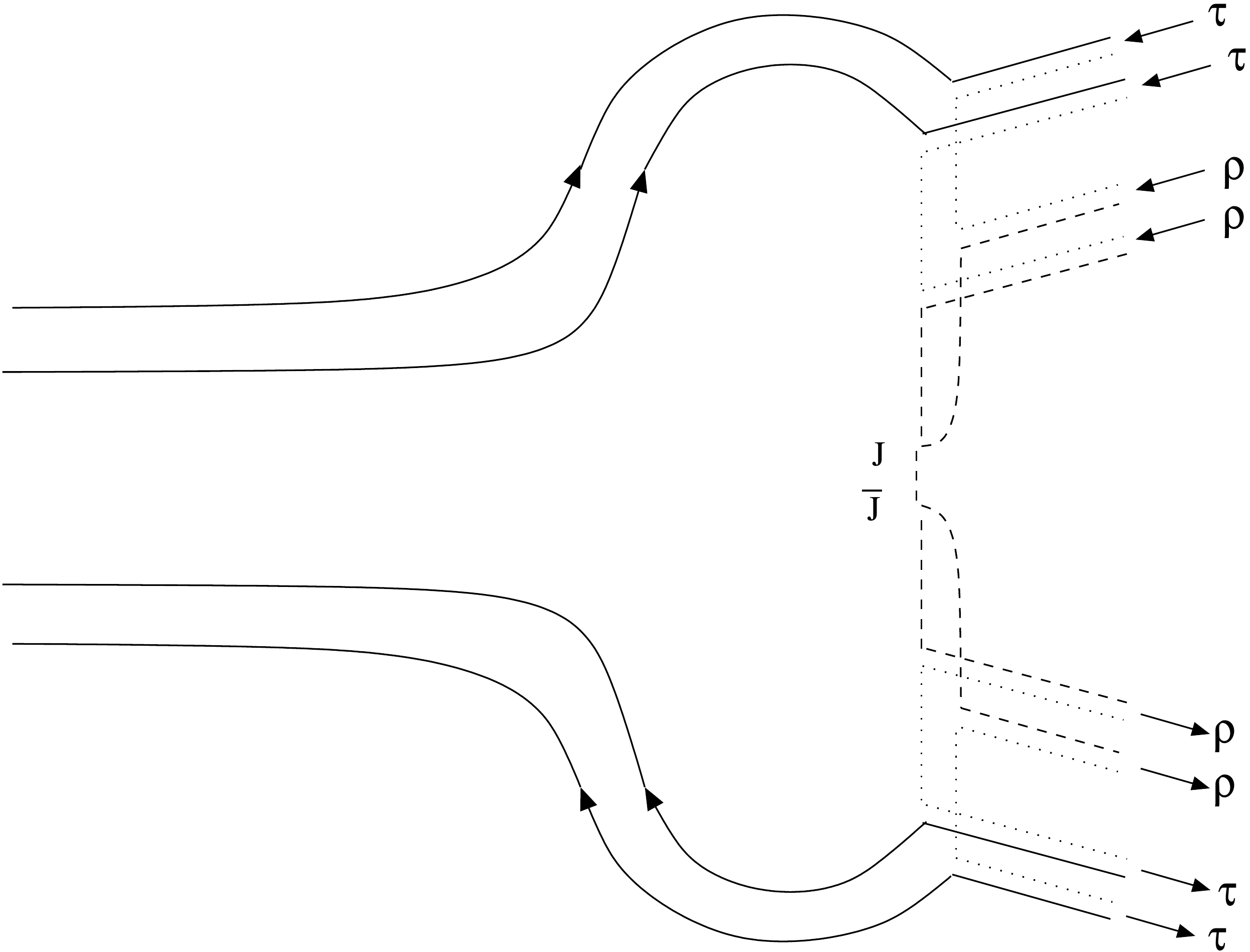}
\caption{Detailed representation of the reduced vertex with external strands.}
\label{fig:fullvertex11}
\end{figure}

The main difference comes from the fact that the strand $1$ appears only as a direct contraction of the $J$ and $\bar J$
corresponding to the same mark, but does not appear anymore among the indices of the resolvents. 
In particular this leads to a novel graphical representation, in which the solid strands of color $1$ have been 
erased, there are no $\tau_1$ or $\rho_1$ ribbon edges, and the $J$ and $\bar J$ of a mark 
are connected by a dashed strand of color $1$. We call such a vertex \emph{reduced} and we picture it
like in Figure \ref{fig:fullvertex11}.

\section{The mixed expansion}\label{sec:mixed}

Let us go back to eq. \eqref{cumulants1},
\begin{align*}
&\mathfrak{K}(\rho_{\cD})=
\sum_{v\geq k}\frac1{v!}\ \frac{(-2\lambda)^{v-1}}{N^{(D-1)(k+v-1) }} \sum_{\tau_{\cD}} \left( \prod_{c = 1}^D \mathrm{Wg}(N,\tau_c \rho_c^{-1})\right) \crcr
&\times \sum_{\mathcal{T}_{v,\{i_d\},\{{\mathcal{C}}(l)\}}}
\int_0^1 \left(\prod_{l_{ij}\in T_v}du_{ij}\right)\int d\mu_{T_v,u}(\sigma) \sum_{n,m} \\
&\times\left(\prod_{i=1}^v\prod_{p=1}^{\mathrm{res}(i)}  R(\sigma^i)_{ n_{i,p} m_{i,p}}\right)
\left(\prod_{l\in T_v}\delta_{\cD  }^{l,{\mathcal{C}}(l)}\right)
\left( \prod_{d=1}^k \prod_{c=1}^D \delta_{n^c_{i_d,q+1}\, m_{i_{\tau_c(d)},q}^c}\right) \nonumber \; ,
\end{align*}
and refine a term in this sum by Taylor expanding up to an order $L$ using the formula
\begin{align*}
f(\sqrt{\lambda})=& \sum_{q=0}^{L}\frac1{q!}\left[\frac{d^q}{dt^q}f(\sqrt{t\lambda})\right]_{t=0}\ \crcr
   & +\ \frac1{L !}\int_0^1 dt \; (1-t)^{L}\frac{d^{L+1} }{dt^{L+1} } \left( f(\sqrt{t\lambda}) \right) 
\end{align*}
on the contributions of the trees $\mathcal{T}_{v,\{i_d\},\{{\mathcal{C}}(l)\}}$, using
\[ 
\frac{d}{dt} R(\sqrt{t} \sigma )_{ n m}=
\frac1{2t} \left( \sum_{\mathcal{C}}\left( \sigma_{ab}^{i\ {\mathcal{C}}}\frac{\dr}{\dr\sigma_{ab}^{i\ {\mathcal{C}}}} 
+ \bar\sigma_{ab}^{i\ {\mathcal{C}}}\frac{\dr}{\dr\bar\sigma_{ab}^{i\ {\mathcal{C}}}}
\right)
\right) R( \sqrt{t} \sigma)_{n m} \; ,
\]
where we dropped the superscript on the indices of $\sigma^{\cC}$, and integrating by parts 
\begin{align*}
&\frac{d}{dt}\int d\mu_{T_v,u}(\sigma)\left(\prod_{i=1}^v\prod_{p=1}^{\mathrm{res}(i)}  R( \sqrt{t}\sigma^i)_{n_{i,p} m_{i,p}}\right)
=\frac{1}{2t}\int d\mu_{T_v,u}(\sigma)\crcr
&\qquad \times \sum_{i=1}^v\sum_{\mathcal{C}} \left(
\sigma_{ab}^{i\ {\mathcal{C}}}\frac{\dr}{\dr\sigma_{ab}^{i\ {\mathcal{C}}}} 
+ \bar\sigma_{ab}^{i\ {\mathcal{C}}}\frac{\dr}{\dr\bar\sigma_{ab}^{i\ {\mathcal{C}}}}
\right)\left(\prod_{i=1}^v\prod_{p=1}^{\mathrm{res}(i)} R( \sqrt{t}\sigma^i)_{n_{i,p} m_{i,p}}   \right) \crcr
&=\frac{1}{2t}\Bigg{[}e^{\sum_{i,j}w_{ij}\sum_{\mathcal{C}}
\left(
\frac{\dr}{\dr\sigma_{ab}^{i\ {\mathcal{C}}}}\frac{\dr}{\dr\bar\sigma_{ab}^{j\ {\mathcal{C}}}}  \right) } \crcr
& \qquad \times 
\sum_{i,\,{\mathcal{C}}}\left(
\sigma_{ab}^{i\ {\mathcal{C}}}\frac{\dr}{\dr\sigma_{ab}^{i\ {\mathcal{C}}}} 
+ \bar\sigma_{ab}^{i\ {\mathcal{C}}}\frac{\dr}{\dr\bar\sigma_{ab}^{i\ {\mathcal{C}}}}
\right)\left(\prod_{i,\,p}  R( \sqrt{t}\sigma^i)_{n_{i,p} m_{i,p}} \right) \Bigg{]}_{\sigma=0} \crcr
&=\Bigg{[}e^{\sum_{i,j}w_{ij}\sum_{\mathcal{C}}\left(
\frac{\dr}{\dr\sigma_{ab}^{i\ {\mathcal{C}}}}\frac{\dr}{\dr\bar\sigma_{ab}^{j\ {\mathcal{C}}}} 
\right)} \crcr
& \quad \times \sum_{i,j,\,{\mathcal{C}}}\frac{w_{ij}}{2t}\left(
\frac{\dr}{\dr\sigma_{ab}^{i\ {\mathcal{C}}}}\frac{\dr}{\dr\bar\sigma_{ab}^{j\ {\mathcal{C}}}} + \frac{\dr}{\dr\sigma_{ab}^{j\ {\mathcal{C}}}}\frac{\dr}{\dr\bar\sigma_{ab}^{i\ {\mathcal{C}}}}
\right)\left(\prod_{i,\,p}   R( \sqrt{t}\sigma^i)_{n_{i,p} m_{i,p}}    \right) \Bigg{]}_{\sigma=0} \crcr
& = \Bigg{[}e^{\sum_{i,j}w_{ij}\sum_{\mathcal{C}}\left(
\frac{\dr}{\dr\sigma_{ab}^{i\ {\mathcal{C}}}}\frac{\dr}{\dr\bar\sigma_{ab}^{j\ {\mathcal{C}}}} 
\right)} \crcr
& \quad \times \sum_{i,j,\,{\mathcal{C}}}\frac{w_{ij}}{t}\left(
\frac{\dr}{\dr\sigma_{ab}^{i\ {\mathcal{C}}}}\frac{\dr}{\dr\bar\sigma_{ab}^{j\ {\mathcal{C}}}} 
\right)\left(\prod_{i,\,p}   R( \sqrt{t}\sigma^i)_{n_{i,p} m_{i,p}}    \right) \Bigg{]}_{\sigma=0} 
\; .
\end{align*}
The sum over $i,j$ and $\cC$ is a sum over all the ways of adding a loop edge to the graph 
$\mathcal{T}_{v,\{i_d\},\{{\mathcal{C}}(l)\}}$. Evaluating the derivatives with respect to $\sigma^{\cC}$ and $\bar \sigma^{\cC}$ we see that the loop edge gives the 
same kind of colored contraction $\delta^{l,{\mathcal{C}}}_{\cD}$ as a tree edge.
Furthermore, each loop edge brings a factor $\frac{-2t\lambda}{N^{D-1}}$ (hence the $t$'s cancel), because the matrix $w_{ij}$ is symmetric
and the same loop edge $ij$ is generated by two terms: $\frac{\dr}{\dr\sigma_{ab}^{i\ {\mathcal{C}}}}\frac{\dr}{\dr\bar\sigma_{ab}^{j\ {\mathcal{C}}}}  $
and $\frac{\dr}{\dr\sigma_{ab}^{j\ {\mathcal{C}}}}\frac{\dr}{\dr\bar\sigma_{ab}^{i\ {\mathcal{C}}}} $.

Repeating this process $L$ times gives a sum of $\frac{(2v+k-3+2L)!}{(2v+k-3)!}$ 
terms labeled by trees $\mathcal{T}_{v,\{i_d\},\{{\mathcal{C}}(l) \} }$ decorated with $L$ colored,
labelled loop edges forming the set $\cL$,
\begin{align*}
&\frac{d^q}{dt^q}
\int_0^1 \left(\prod_{l_{ij}\in T_v}du_{ij}\right)\int d\mu_{T_v,u}(\sigma) 
\left(\prod_{i=1}^v\prod_{p=1}^{\mathrm{res}(i)}  R( \sqrt{t}\sigma^i)_{n_{i,p} m_{i,p}}    \right) \crcr
& \qquad \times \left(\prod_{l\in T_v}\delta_{\cD}^{l,{\mathcal{C}}(l)}\right)
\left( \prod_{d=1}^k \prod_{c=2}^D \delta_{n^c_{i_d,q+1}\, m_{i_{\tau_c(d)},q}^c}\right)\nonumber\\
&=\left(\frac{(-2\lambda)}{N^{D-1}}\right)^q \sum_{ \cL, |\cL|=q }
\int_0^1 \left(\prod_{l_{ij}\in T_v}du_{ij}\right)\int d\mu_{T_v,u}(\sigma) \left(\prod_{l\in \cL }\delta_{\cD }^{l,{\mathcal{C}}(l)} w_{i(l)j(l)}\right)\nonumber\\
&\qquad \times\left(\prod_{i=1}^v\prod_{p=1}^{\mathrm{res}(i)}  R( \sqrt{t}\sigma^i)_{n_{i,p} m_{i,p}}  \right)
\left(\prod_{l\in T_v}\delta_{\cD}^{l,{\mathcal{C}}(l)}\right)
\left( \prod_{d=1}^k \prod_{c=2}^D \delta_{n^c_{i_d,q+1}\, m_{i_{\tau_c(d)},q}^c}\right).
\end{align*}

Taking into account that $R(0) = {\bf 1}^{\cD}$ proves the theorem because 
the first $L$ terms of the Taylor expansion up to order $L$ can be evaluated explicitly: one obtains a free sum for each of the 
internal and external faces of the graph, and theorem \ref{th:mixed} follows.

\section{Absolute convergence}\label{absconv}

In order to establish the absolute convergence of the series in eq. \eqref{eq:mixedexp}, we need to establish a bound on an individual term. 
The explicit terms (consisting in trees with up to $L$ loops) and the rest term (trees with $L+1$ loops) are bounded by very different methods,
explained in the next two subsections.

\subsection{Bounds on the explicit terms}

The global scaling in $N$ of the term associated to the tree $\cT$ decorated with $q$ loop edges $\cL$ and external strands $\tau_{\cD}$
in eq. \eqref{eq:mixedexp} is
\[
 \frac{1}{N^{(D-1)(k+v-1) }} \frac{1}{N^{q(D-1)}} \; \left( \prod_{c =1}^D  
 \frac{1}{N^{2k}} N^{C(\tau_c \rho_c^{-1})} \right) N^{F_{\rm int} (\cT,\cL) + F_{ {\rm ext} } (\cT,\cL,\tau_{\cD})   } \;,
\] 
where we used the asymptotic behavior \eqref{eq:WeinlargeN} of the Weingarten functions. We thus need to bound the number of faces (internal or external)
of the  tree $\cT$ decorated by the loop edges $\cL$.

Recall that $\mathfrak{C}(\rho_{\cD }) $ denotes the number of connected components of the graph associated to the permutations $\rho_{\cD }$.
We denote (naturally) $F_{\rm int} (\cT)$ and  $F_{ {\rm ext} } (\cT,\tau_{\cD})$ the numbers of internal and external faces of the tree $\cT$ itself 
(with external strands $\tau_{\cD}$) with no loop edges.

\begin{lemma}\label{lem:crucial1}
 We have the following bounds:
 \begin{align}\label{eq:goodbound1}
    F_{\rm int} (\cT,\cL) + F_{ {\rm ext} } (\cT,\cL,\tau_{\cD})  
   \le  F_{\rm int} (\cT) + F_{ {\rm ext} }(\cT,\tau_{\cD}) + q  \frac{D}{2}  \; ,
 \end{align}
\begin{equation}\label{eq:goodbound}
   \sum_{c = 1}^D C(\tau_c \rho_c^{-1}) + F^{\cD}_{\rm int}(\cT) + F^{\cD}_{ {\rm ext} }(\cT,\tau_{\cD}) \le (D+1) k + D + (D-1) v - \mathfrak{C}(\rho_{\cD}) \; .
 \end{equation}
 \end{lemma}
Before proving these statements, let us comment on a subtle point: this bound holds for trees having at most a mark per vertex. 
In the next section we will use the Cauchy Schwarz inequalities which lead to vertices having several marks in order to bound the rest term.
The scaling with $N$ of such graphs obeys a weaker bound and, in order to establish theorem \ref{th:conv} we will need to push the expansion up to a 
relatively high (but finite) number of loops.

\noindent{\bf Proof:} Equation \eqref{eq:goodbound1} is trivial, taking into account that for all loop edges $|\cC|\le D/2$ and adding 
an loop edge on a graph can at most divide $|\cC|$ faces into two. 

The proof of eq. \eqref{eq:goodbound} is somewhat more involved. It is done by an iterative procedure consisting in deleting at each step a leaf (univalent vertex) 
of the tree together with the tree edge it is hooked to and tracking the evolution of 
$   \sum_{c = 1}^D C(\tau_c \rho_c^{-1}) + F_{\rm int}(\cT) + F_{ {\rm ext} }(\cT,\tau_{\cD}) + \mathfrak{C}(\rho_{\cD}) $.

Chose the univalent vertex $i$, connected to the rest of the tree by an edge of colors $\cC$. There exists an unique vertex in the tree to 
which $i$ is hooked, called its \emph{ancestor}. The deletion is defined as follows:
\begin{itemize}
 \item if $i$ has no mark, the deletion consists in erasing all the faces with color in $\cD \setminus \cC$ containing $i$ and 
       reconnecting the faces with color in $\cC$ passing through $i$ directly on its ancestor. 
 \item if $i$ has a mark, then it is one of the marked vertices $i_1,\dots i_k$. Say $i$ is $i_d$. 
       The deletion proceeds in two steps. 
       
       If $\tau_c (d) \neq d $ then we replace it by the permutation $\tilde \tau_c$ defined as:
          \[
                  \begin{cases}
          \tilde\tau_c (d) = d  \; , \\
          \tilde \tau_c \left( \tau_c^{-1} (d)\right) =  \tau_c(d) \; , \\
          \tilde \tau_c (d') = \tau_c(d') \; , \quad \forall d' \neq d, \tau_c^{-1}(d) .
                                                       \end{cases} 
          \]
         Graphically this comes to cutting the ribbon edges representing the permutations $\tau_c$ incident at $i_d$ 
         and reconnecting them the other way around. 
       
         Similarly, if $\rho_c (d) \neq d $ then we replace it by the permutation $\tilde \rho_c$ defined as:
          \[
          \begin{cases}
          \tilde\rho_c (d) = d  \; , \\
          \tilde \rho_c \left( \rho_c^{-1} (d)\right) =  \rho_c(d) \; , \\
          \tilde \rho_c (d') = \rho_c(d') \; , \quad \forall d' \neq d, \rho_c^{-1}(d) .
                                                       \end{cases} 
          \]
         Graphically this comes to cutting the ribbon edges representing the permutations $\rho_c$ incident at $i_d$ 
         and reconnecting them the other way around. 
    
        For a vertex $i_d$ such that $\tau_c(d) = d$ and $\rho_c(d) = d$ for all colors, the deletion consists in reconnecting the 
        solid strands of color in $\cC$ on its ancestor and deleting all the other strands (solid, dashed and dotted).
\end{itemize}

\begin{figure}[ht]
\centering
\includegraphics[height=4.5cm]{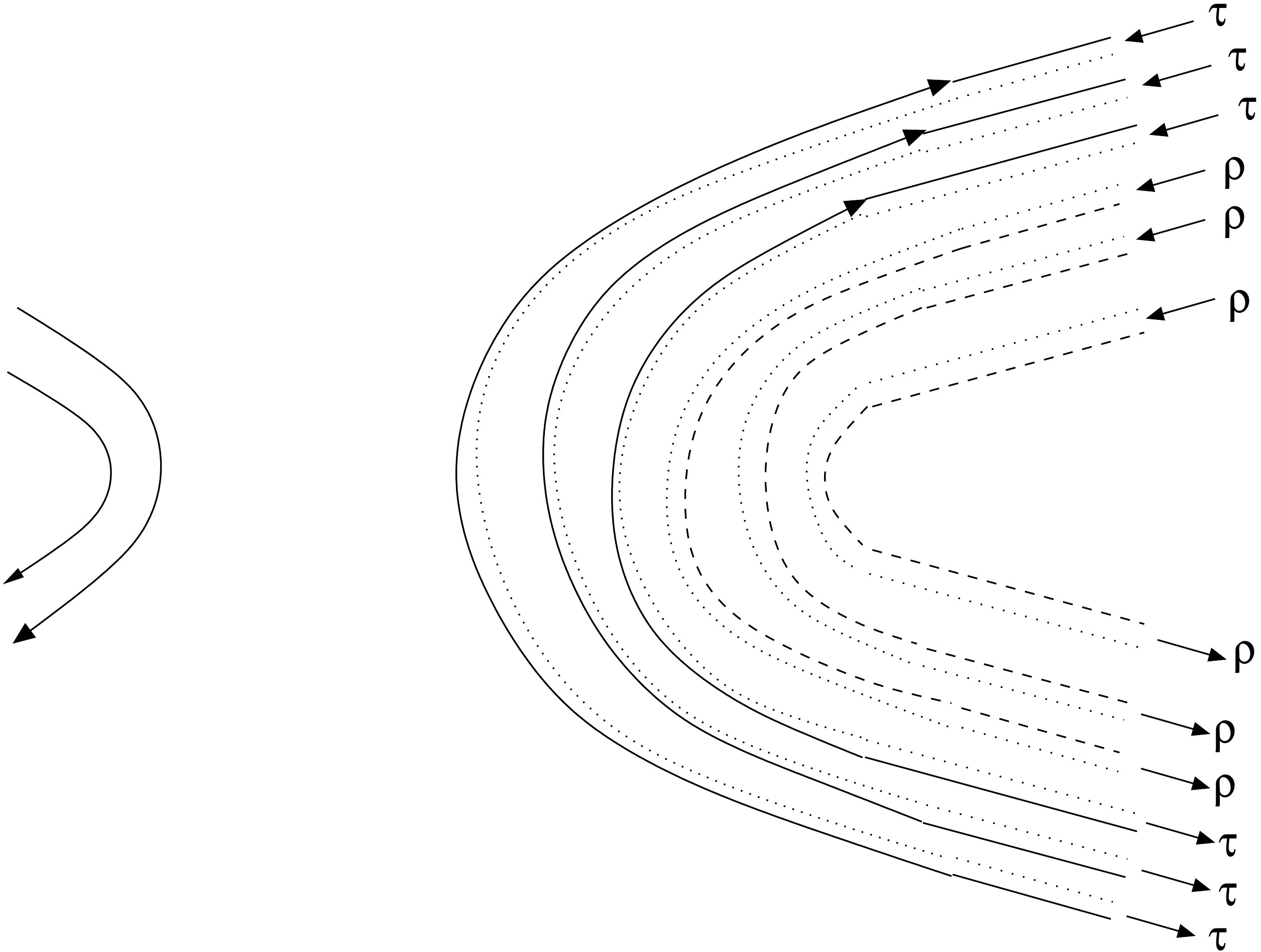}
\caption{Deletion of a vertex with external sources.}
\label{fig:fullvertex1}
\end{figure}

       For a vertex $i_d$ (represented in Figure \ref{fig:fullvertex}) such that for all colors $\tau_c(d)\neq d$ and $\rho_c(d)\neq d$, 
        the deletion leads to the drawing in Figure \ref{fig:fullvertex1}. 

Let us denote $ \tau_c^r$, $\rho_c^r$ and $\cT^r$ the permutations and the tree obtained after having erased one vertex. If the vertex had a mark and the deletion 
was done in two steps, at the intermediary step we have the same tree $\cT$, but the permutations $\tilde \tau_{\cD}$ and $\tilde \rho_{\cD}$. 

If $i$ has no mark, the permutations $\tau$ and $\rho$ are unaffected by the deletion. As we erase at most $|\cD\setminus \cC|\le D-1$ faces,
we have
\begin{align*}
&  \sum_{c = 1}^D C(\tau_c \rho_c^{-1}) + F_{\rm int}(\cT) + F_{ {\rm ext} }(\cT,\tau_{\cD})  + \mathfrak{C}(\rho_{\cD}) \le \crcr
& \qquad  \le   \sum_{c = 1}^D C(\tau^r_c [\rho_c^r]^{-1}) + F_{\rm int}(\cT^r) + F _{ {\rm ext} }(\cT^r,\tau_{\cD}^r) + \mathfrak{C}(\rho^r_{\cD}) +D-1  \; .
\end{align*}
and the graph obtained after deletion has one fewer vertex.

We now analyze the evolution of this quantity in the case when $i$ has a mark, $i=i_d$ and we go through the intermediary graph:
\begin{itemize}
 \item \emph{The number of connected components of the graph of the permutations $\rho_{\cD}$}. 
        We always create a connected component with a black and a white vertex $d$ and edges $\tilde \rho_c(d) =d$. The other connected components 
       of the graph associated to $\rho_{\cD}$ change as follows:
       \begin{itemize}
        \item if $\rho_c(d) \neq d$ for all $c \in \cD$ then there are two cases:
               \begin{itemize}
                \item the black and white vertices $d$ belong to the same connected component. Then by going to $\tilde \rho_{\cD}$
                      this components either survives or it is split into several connected components. 
                      Thus $ \mathfrak{C}(\rho_{\cD})  \le \mathfrak{C}( \tilde \rho_{\cD}) -1 $.
                \item the black and white vertices $d$ belong to two distinct connected components. Then the two are either merged into 
                 an unique component, or split into several. We always have $ \mathfrak{C}(\rho_{\cD})  \le \mathfrak{C}( \tilde \rho_{\cD})$. 
               \end{itemize}
        \item if there exists $c\in \cD$ such that $ \rho_c(d) = d$.  Then the black and white vertex $d$ belong to the same connected component
              and this component either survives or it is split into several by going to $\tilde \rho_{\cD}$. Hence in this case we always 
              have $ \mathfrak{C}(\rho_{\cD})  \le \mathfrak{C}( \tilde \rho_{\cD}) -1  $.
       \end{itemize}
          To summarize:
             \[
              \begin{cases}
                \mathfrak{C}(\rho_{\cD})  \le \mathfrak{C}( \tilde \rho_{\cD}) -1  \text{ if } \exists c\;,\;\; \rho_c(d) = d \; , \\
                \mathfrak{C}(\rho_{\cD})  \le \mathfrak{C}( \tilde \rho_{\cD}) \text{ otherwise} \; .
              \end{cases} 
             \]
 \item \emph{The number of faces of color $c\in \cC$.} This number is:
         \begin{itemize}
         \item constant if $\tau_c(d)=d$.
         \item can at most decrease by $1$ if $\tau_c(d) \neq d$.
         \end{itemize}
       The number of faces of color $c\in \cD \setminus \cC$ is:
         \begin{itemize}
         \item constant if $\tau_c(d)=d$.
         \item increases by $1$ if $\tau_c(d) \neq d$.
         \end{itemize}
         Denoting $s^{\cC} \le |\cC|$ the number of colors in $\cC$ such that $\tau_c(d) \neq d$, and $s^{\cD\setminus \cC}$ the number of colors in 
         $\cD \setminus \cC$ such that $\tau_c(d) \neq d$, we get: 
    \[
         F _{\rm int}(\cT) +  F _{ {\rm ext} }(\cT,\tau_{\cD} ) \le  F_{\rm int} (\cT) +  F_{ {\rm ext} } (\cT,\tilde \tau_{\cD})  + s^{\cC} - s^{\cD \setminus \cC}  \; .
    \]
 \item \emph{The number of faces $\tau_c\rho_c^{-1}$.} This number changes as follows: 
         \begin{itemize}
           \item   if $\tau_c(d) = d$,
          \begin{itemize}
           \item if $\rho_c(d) = d$, $C(\tau_c \rho_c^{-1})$ is constant.
           \item if $\rho_c(d) \neq d$, $C(\tau_c \rho_c^{-1})$ increases by $1$.
          \end{itemize}
        
          \item if $\tau_c(d) \neq d$, \begin{itemize}
                                         \item if $\rho_c(d) = d$, $C(\tau_c \rho_c^{-1})$ increases by $1$
                                         \item if $\rho_c(d) \neq d $, $C(\tau_c \rho_c^{-1})$ can not decrease.
                                       \end{itemize}
         Hence:
            \[
              \begin{cases}
                  \sum_{c = 1}^D C(\tau_c \rho_c^{-1})  \le  \sum_{c = 1}^D C(\tilde \tau_c \tilde \rho_c^{-1}) -1 
                     \text{ if } \exists c  \begin{cases}
                                             \tau_c(d) = d \\ \text{ and } \\ \rho_c(d) \neq d 
                                             \end{cases}  \; ,  \\
                  \sum_{c = 1}^D C(\tau_c \rho_c^{-1})  \le  \sum_{c = 1}^D C(\tilde \tau_c \tilde \rho_c^{-1}) \text{ otherwise} \; .
              \end{cases}
             \]
       \end{itemize}
\end{itemize}      

    Remark that $s^{\cC} -s^{\cD\setminus \cC} \le |\cC| -1$ unless both $s^{\cC} = |\cC|$ and $s^{\cD \setminus \cC}=0$. However in this case
    $\tau_c(d) = d$ for all $c\in \cD\setminus \cC$, and either $ \rho_c(d) \neq d $ or $\rho_c(d) = d$ hence in all cases we obtain the bound:
    \begin{align}\label{eq:problem}
   &  \sum_{c = 1}^D C(\tau_c \rho_c^{-1}) + F _{\rm int}(\cT) + F _{ {\rm ext} } (\cT ,\tau_{\cD}) + \mathfrak{C}(\rho_{\cD}) \le \crcr
 & \qquad \le \sum_{c = 1}^D C(\tilde \tau_c \tilde \rho_c^{-1}) +  F _{\rm int}(\cT) +  F_{ {\rm ext} } (\cT,\tilde \tau_{\cD})
 + \mathfrak{C}(\tilde \rho_{\cD}) + |\cC| -1 \; .
    \end{align}

    Finally, erasing a vertex $i_d$ with $\tau_c(d) = d$ and $\rho_c(d)=d$ for all $c$ we obtain
 \begin{align*}
 &  \mathfrak{C}( \tilde \rho_{\cD}) =  \mathfrak{C}(   \rho^r_{\cD}) +1 \qquad 
    C(\tilde \tau_c \tilde \rho_c^{-1})  =    1 +  C(\tau^r_c [\rho^r_c]^{-1})  \crcr
 &  \tilde F_{\rm int} (\cT) +  F_{ {\rm ext} } (\cT,\tilde \tau_{\cD}) =  F _{\rm int}(\cT^r) +  F_{ {\rm ext} } (\cT^r,\tau^r_{\cD}) + |\cD \setminus \cC| \; ,
\end{align*}
hence
\begin{align*}
 & \sum_{c = 1}^D C(\tau_c \rho_c^{-1}) + F_{\rm int}(\cT) + F _{ {\rm ext} } (\cT,\tau_{\cD}) + \mathfrak{C}(\rho_{\cD}) \le 
  \sum_{c=1}^D  C(\tau^r_c [\rho^r_c]^{-1}) 
 \crcr
& \qquad   +  F _{\rm int}(\cT^r) +  F_{ {\rm ext} } (\cT^r,\tau^r_{\cD})  +  \mathfrak{C}(   \rho^r_{\cD}) +
D + |\cD \setminus \cC| + |\cC|  \; ,
\end{align*}
and the number of vertices and of marks both go down by $1$. Iterating up to the last vertex, we either end up with a vertex with no mark or with a vertex 
with a mark, and all $\rho_c (d) = d ,\tau_c (d) =d$. Counting the faces and the number of connected components of $\rho_{\cD}$ of the two possible 
end graphs we obtain:
\begin{align*}
& \sum_{c = 1}^D C(\tau_c \rho_c^{-1}) + F _{\rm int}(\cT) + F _{ {\rm ext} }(\cT,\tau_{\cD}) + \mathfrak{C}(\rho_{\cD}) \le \crcr
& \qquad \le  \begin{cases}
    k(D+1) + (D-1)(v-1) + D  \; , \\
    (k-1)(D+1) + (D-1) (v-1) + 2D +1 \; ,
 \end{cases}
\end{align*}
which proves eq. \eqref{eq:goodbound}.

The main difficulty in the above proof is to obtain the $-1$ on the right hand side of \eqref{eq:problem}. This relies crucially on the observation that 
in a tree with a mark per vertex, when $s^{\cD \setminus \cC}=0$, then the strands with colors $c \in \cD \setminus \cC$ are such that 
$\tau_c(d) = d$. This is fine as long as we have a unique mark per vertex, however if we have several marks, the proof fails.

\qed

\subsection{Bounds on the rest term}

In order to establish a bound on the rest terms
in equation \eqref{eq:mixedexp} we will use the technique introduced in \cite{MagRiv} and \cite{Magnen:2009at} of iterated Cauchy-Schwarz inequalities:
\begin{align}\label{cauchy}
 | \bra{A} R\otimes R'\otimes \mathbf{1}^{\otimes p}\ket B | \;   \leq  \;  \|R\|\|R'\|\sqrt{\braket {A|A}}\sqrt{\braket{B|B}} \; .
\end{align}

\begin{lemma}\label{lem:boundrest}
We have the bound:
\begin{align}\label{eq:rest}
& \Bigg{|}\int d\mu_{T_v,u}(\sigma) \sum_{m,n} \left(\prod_{i=1}^v\prod_{p=1}^{\mathrm{res}(i)}  R( \sqrt{t}\sigma^i)_{n_{i,p} m_{i,p}}  \right)
 \left(\prod_{l\in \cL }\delta_{\cD }^{l,{\mathcal{C}}(l)} \right)  
\crcr
&\qquad \times \left(\prod_{l\in T_v}\delta_{\cD}^{l,{\mathcal{C}}(l)}\right)
\left( \prod_{d=1}^k \prod_{c=2}^D \delta_{n^c_{i_d,q+1}\, m_{i_{\tau_c(d)},q}^c}\right) \Bigg{|} \le \crcr
& \le  N^{D + Dk + (D-1)(v-1)+    |\cL| \frac{D}{2}}
 \; .
\end{align} 
\end{lemma}

\noindent{\bf Proof:} The rest term is a contraction of resolvents placed around the vertices in a pattern encoded in a tree 
with loop edges and external strands.
On any tree, one can choose a resolvent and then count the resolvents following the clockwise contour walk of the tree, indexing 
them from $R_1$ to $R_{2n}$ (or $R_{2n+1}$). 

\begin{figure}[ht]
\centering
\includegraphics[height=4.5cm]{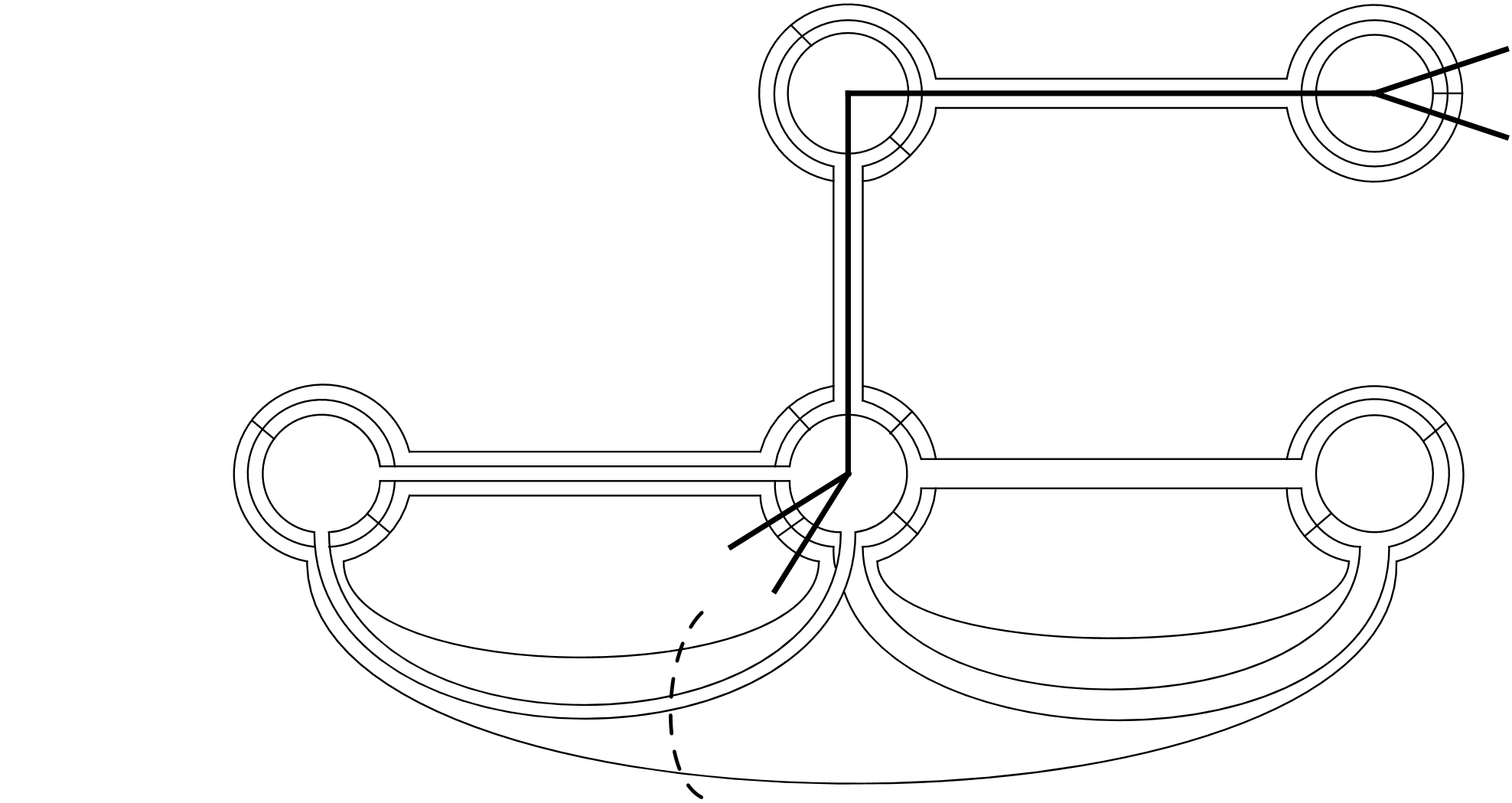}
\caption{Splitting the graph in two part in order to apply the Cauchy-Schwartz inequality. 
 }
\label{graph2'}
\end{figure}

Choosing $R_1$ and $R_{n+1}$ as the $R$ and $R'$ of formula (\ref{cauchy}), the  
vector $A$ is made of all the resolvents from $R_2$ to $R_n$ and the contractions between them.
The vector $B$ is made of all the resolvents from $R_{n+2}$ to $R_{2n}$ (or $R_{2n+1}$) and the contractions between them.
The contractions of indices of the resolvents $R_2$ to $R_n$ with indices of the resolvents $R_{n+2}$ to $R_{2n}$ (or $R_{2n+1}$),
which can exist due to the $\tau$ external strands or due to the loop edges, is encoded in the $\mathbf{1}^{\otimes p}$ operator.
If an index of $R_1$ is directly contracted with an index of $R_{n+1}$, the latter contributes a Kronecker delta to the vector $A$ or $B$.
We forget the dashed and dotted strands, as they do not contribute to the quantity on the left hand side of equation \eqref{eq:rest}.
We represented such a splitting in Figure \ref{graph2'}.

We apply the formula (\ref{cauchy}) and, because the norm of the resolvent is bounded by $1$, $\|R(\sigma)\|\leq 1$, 
\begin{align}
 | \bra{A} R_1\otimes R_{n+1}\otimes \mathbf{1}^{\otimes p}\ket B | \; \leq \; \sqrt{\braket {A|A}\braket{B|B}}.
\end{align}
The resolvents $R_1$ and $R_{n+1}$ were on the vertices of the tree
and the previous splitting in $A$ and $B$ corresponds to cutting through the interior of the tree (the cut is represented in bold 
in Figure \ref{graph2'}). 
Any contraction strand between the part $A$ and the part $B$ (due to the external strands $\tau$ or to the loop edges) is also cut.

The scalar products $\braket {A|A}$ and $\braket{B|B}$ are half graphs merged with their mirror symmetric with respect to the splitting line.
They also have the structure of trees with loop edges and external $\tau$ strands. However, in contrast 
with the original graph, they can have several marks on the same vertex. This happens whenever the resolvent 
$R_1$ (or $R_n$) belongs to a marked vertex. This is ultimately the reason for which the bound on the rest term we establish below is weaker than
the bound on the explicit terms of lemma \ref{lem:crucial1}.

\begin{figure}[ht]
\centering
\begin{tabular}{c}
\includegraphics[height=4.3cm]{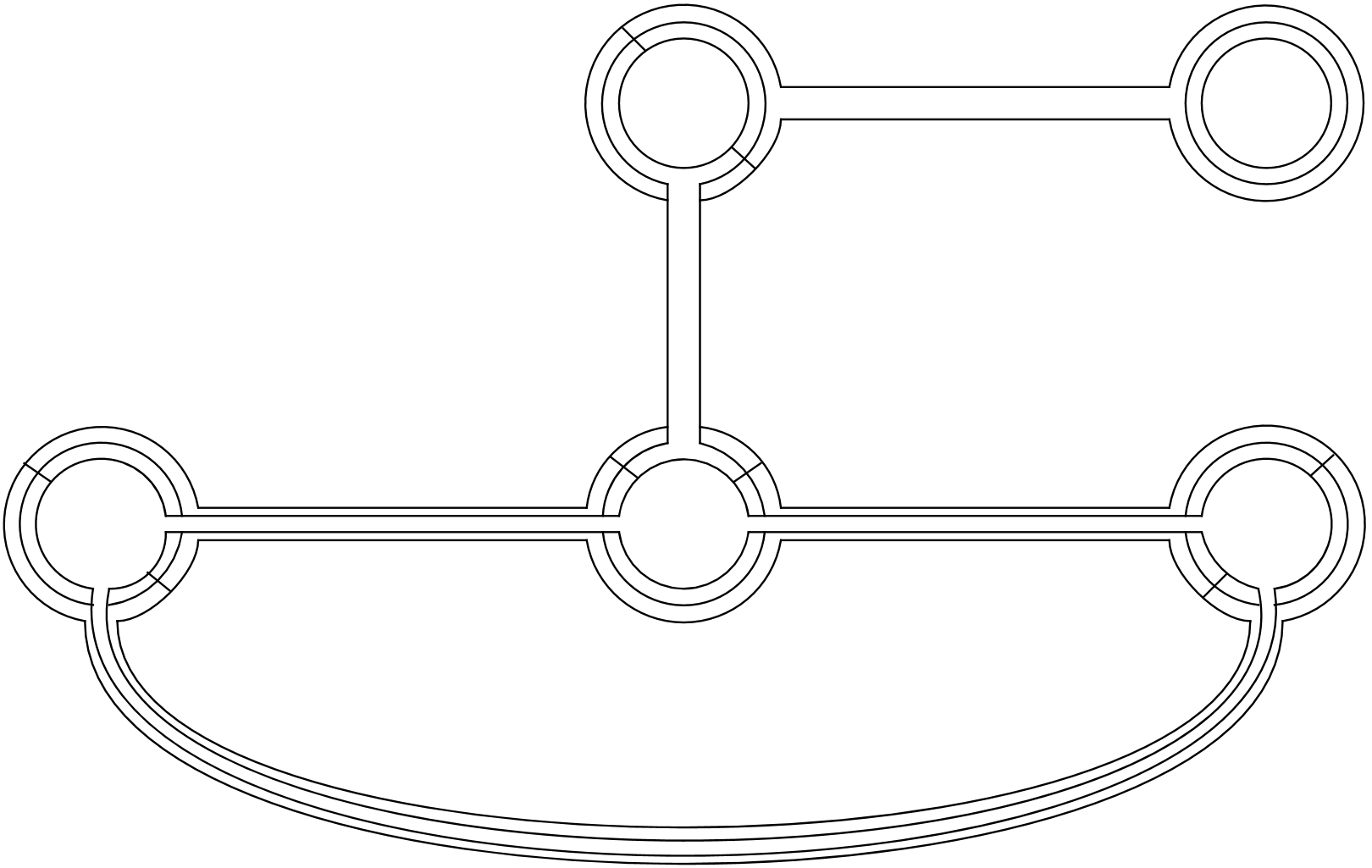} \\ \hspace{.5cm} \\ \includegraphics[height=5cm]{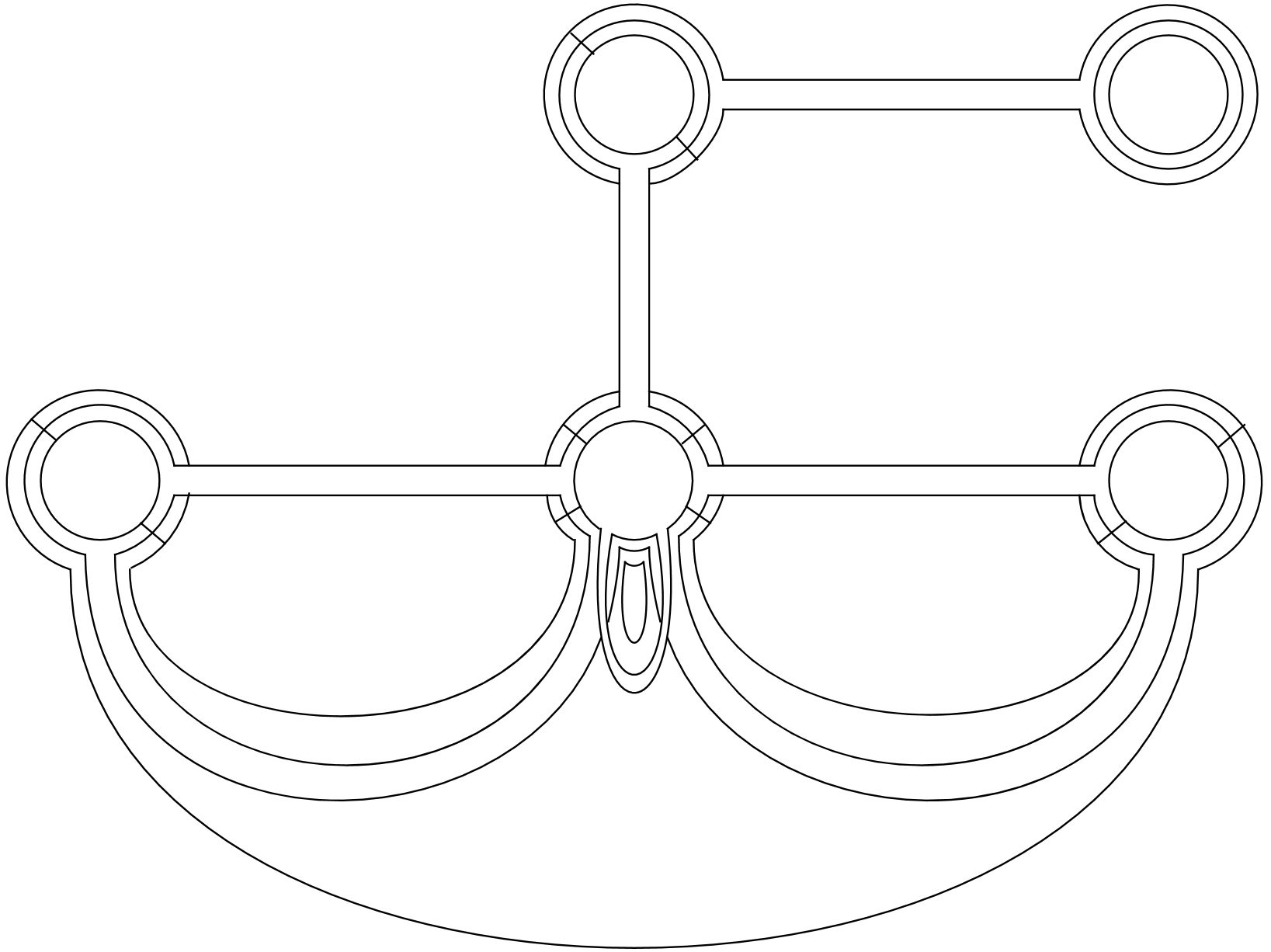} \end{tabular}
\caption{The scalar product graphs $\braket {A|A}$ (top) and $\braket{B|B}$ (bottom). }
\label{graph2ab}
\end{figure}

As the resolvents $R_1$ and $R_{n+1}$ have been taken out, the scars (i.e. the corners where $R_1$ and $R_{n+1}$ were connecting on $A$ and $B$)
are now just direct identifications of indices. That is, in the graphs merged with their mirror symmetric,
two resolvents have been set to the identity ${\bf 1}^{\cD}$ operator. 
The corresponding corners will be represented as just $D$ parallel strands, with 
no vertical line, like in Figure \ref{graph2ab}.

Let us denote $v^A$, $k^A$, $\cT^{A}$, $\cL^{A}$, $\tau_{\cD}^A$ the number of vertices, marked vertices,
the tree, the loop edges and the permutations of external strands corresponding to the graph $\braket{A|A}$ and similarly for $B$. 
We have the following (in)equalities:
\begin{itemize}
 \item the number of vertices doubles
           \[ 2v = v^A + v^B  \; .\]
 \item the number of marked vertices doubles 
           \[ 2 k = k^A + k^B \; .\]
 \item the number of loop edges doubles
           \[ |\cL| = |\cL^A| + |\cL^B| \; . \] 
 \item the number of faces (closed solid strands) \emph{at least doubles}
          \begin{align*} 
          2F_{\rm int}(\cT, \cL) + 2F_{\rm ext} (\cT,\cL,\tau_{\cD}) & \le F_{\rm int}(\cT^A, \cL^A) + F_{\rm ext} (\cT^A,\cL^A,\tau^A_{\cD}) \crcr
          & 
            +  F_{\rm int}(\cT^B, \cL^B) + F_{\rm ext} (\cT^B,\cL,\tau^B_{\cD}) \; .
         \end{align*}
         This is because a face is either untouched by the splitting line, or it is cut in two pieces (and in both cases it leads to two faces in 
           the mirrored graphs), or it is cut in at least four pieces in which case it leads to at least four faces in the mirrored graphs.
\end{itemize}

Each mirrored graph has an even number of resolvents and contributes to the bound by the square root of its amplitude. 
We can now iterate the process, each time eliminating two resolvents, until there are no resolvents left on any graph. 

If we start with $2n$ resolvents we obtain $2^n$ final graphs. If we start with $2n+1$ resolvents, we add an iteration in which all but one of the resolvents
are already set to the identity operator. The $2^n$ (resp. $2^{n+1}$) final graphs we obtain are made solely 
of faces which represent traces of identity, hence  each  such face brings a factor $N$. 

We denote for, $q=1,\dots 2^n$ (or $2^{n+1}$) by $k^q$, $v^q$, $\cT^q$, $\cL^q$ the numbers of marked vertices, vertices, etc. of the final graphs.
As before, adding a loop edge with color $\cC$ to a final graph can at most divide $|\cC|$ of its faces, hence
\[
 F_{\rm int}(\cT^q, \cL^q) +  F_{\rm ext}(\cT^q, \cL^q,\tau_{\cD}^q) \le  F_{\rm int}(\cT^q) +  F_{\rm ext}(\cT^q,\tau_{\cD}^q) + |\cL^q|\frac{D}{2} \;.
\]
A tree $\cT^q$ with multiple marks on the same vertex has exactly $2Dk^q$ 
ends of strands connected by the permutation $\tau^q$, hence it has at most $Dk^q$ external faces. 
The number of internal faces of $T^q$ is at most $1 + (D-1) v^q $. If we start with $2n$ resolvents we get the bound
\begin{align*}
 & \Bigg{|}\int d\mu_{T_v,u}(\sigma) \sum_{m,n} \left(\prod_{i=1}^v\prod_{p=1}^{\mathrm{res}(i)}  R( \sqrt{t}\sigma^i)_{n_{i,p} m_{i,p}}  \right)
 \left(\prod_{l\in \cL }\delta_{\cD }^{l,{\mathcal{C}}(l)} \right)  
\crcr
&\qquad \times \left(\prod_{l\in T_v}\delta_{\cD}^{l,{\mathcal{C}}(l)}\right)
\left( \prod_{d=1}^k \prod_{c=2}^D \delta_{n^c_{i_d,q+1}\, m_{i_{\tau_c(d)},q}^c}\right) \Bigg{|} \le \crcr
& \le N^{\frac{1}{2^n} \sum_{q=1}^{2^n} \left(  F_{\rm int}(\cT^q, \cL^q) +  F_{\rm ext}(\cT^q, \cL^q,\tau_{\cD}^q) \right) }
\le N^{\frac{1}{2^n} \sum_{q=1}^{2^n} \left( Dk^q + 1 + (D-1)v^q      + |\cL^q|\frac{D}{2}   \right) } \crcr
& = N^{D + Dk + (D-1)(v-1)+    |\cL| \frac{D}{2}}
\end{align*}
and the same holds if we start with $2n+1$ resolvents.

\qed

We are now ready to prove theorem \ref{th:conv}. Taking absolute values in eq. \eqref{eq:mixedexp}, using eq. \eqref{eq:WeinlargeN} 
and using the lemmas \ref{lem:crucial1} and \ref{lem:boundrest} we find:
\begin{align*}
&|\mathfrak{K}(\rho_{\cD})|\leq
\sum_{v\geq k}\frac1{v!}\ \frac{ |2\lambda|^{v-1}}{N^{(D-1)(k+v-1) }} \sum_{\tau_{\cD}} 
\left( \prod_{c = 1}^D \frac{2^{2k}}{N^{2k}} N^{C(\tau_c \rho_c^{-1})}  \right)   \sum_{\mathcal{T}_{v,\{i_d\},\{{\mathcal{C}}(l)\}}} 
   \crcr
&\times
  \Bigg{[}  \sum_{q=0}^L \left(\frac{ | 2\lambda| }{N^{D-1}}\right)^q  \frac{1}{q!} \sum_{\cL, |\cL| = q}
  N^{F_{\rm int} (\cT,\cL )+F_{ \rm ext } (\cT,\cL,\tau_{ \cD} ) } \crcr
& \qquad + \left(\frac{|2\lambda|}{N^{D-1}}\right)^{L+1}  \frac{1}{L!}  \sum_{ \cL, |\cL| = L+1 }
N^{D + Dk + (D-1)(v-1)+    |\cL| \frac{D}{2}} \Bigg{]}   \le \crcr
& \le 2^{2Dk} k!^D \sum_{v\geq k}\frac1{v!} \sum_{\mathcal{T}_{v,\{i_d\},\{{\mathcal{C}}(l)\}}}  \crcr
& \qquad \Bigg{[}  \sum_{q=0}^L  \frac{   | 2\lambda|   ^{v-1+q}  }{q!} \sum_{\cL, |\cL| = q} N^{D - 2(D-1)k - \mathfrak{C}(\rho_{\cD}) 
   - q\left(\frac{D}{2}-1 \right) } \crcr
& \qquad \; + \frac{  | 2\lambda|   ^{v+L}  }{L!} \sum_{\cL, |\cL| = L+1} N^{D - (D-1)k - (L+1) \left(\frac{D}{2}-1 \right)   } \Bigr] \; .
\end{align*}
As $ \mathfrak{C}(\rho_{\cD}) \le k $, choosing $L \ge \frac{Dk}{D/2-1}-1$ ensures that 
\[
 D - (D-1)k - (L+1) \left(\frac{D}{2}-1 \right)  \le D - 2(D-1)k - \mathfrak{C}(\rho_{\cD}) \; ,
\]
and 
\begin{align*}
 |\mathfrak{K}(\rho_{\cD})| & \le  N^{D - 2(D-1)k - \mathfrak{C}(\rho_{\cD}) } \crcr
&    \times   2^{2Dk} k!^D\sum_{v\geq k}\frac1{v!} \sum_{\mathcal{T}_{v,\{i_d\},\{{\mathcal{C}}(l)\}}}  
   \sum_{q=0}^{L+1}  \frac{   | 2\lambda|  ^{v-1+q}  }{q!} \frac{(2v+k-3+2q)!}{(2v+k-3)!} \; .
\end{align*}
Taking into account that 
\begin{align*}
  & \sum_{\mathcal{T}_{v,\{i_d\},\{{\mathcal{C}}(l)\}}}1 \crcr
  &=  (\mathcal N_{\mathcal{Q}})^{v-1}  
   \sum_{d_1 \dots d_v\ge 1 }^{\sum d_i=2(v-1)}\left(\frac{(v-2)!}{\prod_i (d_i-1)!} \prod_i (d_i-1)!\ \times \sum_{i_1,\dots i_k}^{i_k\neq i_{k'}} 
      d_{i_1}...d_{i_k}\right) \crcr
  &=(\mathcal N_{\mathcal{Q}})^{v-1}\frac{v!(2v+k-3)!}{k!(v-k)!(v+k-1)!}\; , 
\end{align*}
with $(\mathcal N_{\mathcal{Q}})^{v-1}$ the number of edge coloring, $d_i$ the degree of the vertex $i$,
$\frac{(v-2)!}{\prod_i (d_i-1)!}$ the number of trees with fixed degrees and $\prod_i (d_i-1)!$ the number of associated plane
trees, and $\prod_{d=1}^k d_{i_d}$ the number of ways to put marks on the vertices $i_d$, we obtain
\begin{align}\label{eq:boundseries}
 |\mathfrak{K}(\rho_{\cD})| & \le  N^{D - 2(D-1)k - \mathfrak{C}(\rho_{\cD}) } \crcr
&    \times   2^{2Dk} k!^{D-1} \sum_{q=0}^{L+1} \sum_{v\geq k} |2\mathcal N_{\mathcal{Q}}\lambda|^{v+q-1}
\frac{(2v+k-3+2q)!}{q!   (v-k)!(v+k-1)!} \; ,
\end{align}
and the sum over $v$ converges for $|8\mathcal N_{\mathcal{Q}}\lambda|<1$.

\subsection*{Second cumulant}

For $k=1$, the rescaled second cumulant can be written
\begin{align*}
 N^{D-1} \mathfrak{K}(\{1\}_{\mathcal{D}})= \sum_{v\geq1} a_v(N,\lambda) ,
\end{align*}
where $a_v$ contains all terms corresponding to graphs with $v$ vertices and is bounded by 
\begin{align*}
  |a_v(N,\lambda)| \leq
  2^{2Dk} k!^{D-1} \sum_{q=0}^{L+1} |2\mathcal N_{\mathcal{Q}}\lambda|^{v+q-1}
\frac{(2v-2+2q)!}{q! v!  (v-1)!}.
\end{align*}
This bound does not depend on $N$ and assures the uniform convergence of the series.

$A_v$ is composed of the amplitude of trees, trees decorated with loops and rest terms. Choosing $L>\frac{Dk}{D/2-1}-1$ ensures the rest terms to be dominated by $1/N$.
A tree decorated with $q$ loops being dominated by $N^{-q(D/2-1)}$, the large $N$ limit of $a_v$ is given by the amplitude of trees.
Moreover, we have $F_{\rm int}(\mathcal{T})+F_{\rm ext}(\mathcal{T},\{1\}_{\mathcal{D}})=(D-1)(v-1)+D$ if and only if $\mathcal{T}$ is composed only of edges with $|\mathcal{C}(l)|=1$.

Thus, the large $N$ limit of the rescaled second cumulant is finite, and is given by the sum over all trees with edges carrying only one color. 
If we denote $\mathcal{N}_1$ the number of interaction terms with $|\mathcal C|=1$, we have

\begin{align*}
&\lim_{N\to\infty} N^{D-1} \mathfrak{K}(\{1\}_{\mathcal{D}}) \ =\  \sum_{v\geq1} \frac{1}{v!} \sum_{\mathcal{T}_{v,i_1,\{\mathcal C(l)\}}^{|\mathcal{C}(l)|=1}}1 \crcr
&= \sum_{v\geq1} (-2\mathcal{N}_1\lambda)^{v-1} \frac{(2v-2)!} {(v-1)!v!} =\frac{-1+\sqrt{1+8\mathcal{N}_1\lambda}}{4\mathcal{N}_1\lambda} .
\end{align*}

This, together with the bound in Eq. \ref{eq:boundseries}, establishes theorem \ref{th:conv}, 
shows that the measure is properly uniformly bounded and thus obeys the universality theorem. 

\section{Uniform Borel summability}\label{BOREL}
A function $f(\lambda, N)$ is said to be Borel summable in $\lambda$ uniformly in $N$ if $f$ is analytic in $\lambda$ in a 
disk $\mathrm{Re}\frac1\lambda>\frac1R$ with $R>0$ independent on $N$ and admits a Taylor expansion
\begin{align*}
 f(\lambda,N)=\sum_{k<r} A_k(N)\lambda^k \ +\ R_{r}(\lambda,N) \; , \;\;  |R_{r}(\lambda,N)|\leq r!  \; a^r  |\lambda|^r K(N) .
\end{align*}
for some $a$ independent of $N$.

\subsection{Analyticity}\label{subsecANA}

To establish the convergence of the series in eq. \eqref{eq:mixedexp} in the domain in the complex plane $\lambda=r e^{i\phi}$, $\phi\in (-\pi,\pi)$
defined by $|\lambda|< \frac{1}{8(\mathcal N_{\mathcal{Q}})} \left( \cos\frac{\phi}{2} \right)^2 $ it is enough to follow step by step the proof of 
theorem \ref{th:conv} and note that the norm of the resolvent is bounded by
\[
  \|R(\sigma)\| \leq \frac{1}{\mathrm{cos}\frac\phi2} \; .
\]
The iterated Cauchy-Schwarz inequalities go through, and it is easy to see that the norm of each resolvent  
contributes to the power 1 to the amplitude of the graph. The total number of resolvents of a graph 
with $v$ vertices and $k$ marks is $2(v-1)+k$. Therefore each term of the overall bound in eq. \ref{eq:boundseries}
must be multiplied by $\left(\frac{1}{\mathrm{cos}\frac\phi2}\right)^{2(v-1)+k}$, which proves the convergence and eq.
\eqref{eq:Borelbound}.

The convergence domain 
 \begin{align*}
  |\lambda|<\frac1{8\mathcal N_{\mathcal{Q}}}\left(\mathrm{cos}\frac\phi2\right)^2 \; ,
 \end{align*}
contains a disk $\mathrm{Re}\frac1\lambda>\frac1R$. In this domain the cumulants eq. \eqref{cumulants0} 
are analytic as the resolvents themselves $R(\sigma^i)_{n_{i,p} m_{i,p}}$ are, which can be proved by verifying the Cauchy-Riemann equation, 
\begin{align*}
r\frac{\dr}{\dr r} R(\sigma^i)_{n_{i,p} m_{i,p}}
=-\frac12 \left[ A(\sigma^i) R(\sigma^i)\right]_{n_{i,p} m_{i,p}} = i\frac{\dr}{\dr\phi} R(\sigma^i)_{n_{i,p} m^{\cD}_{i,p}} \; .
\end{align*}

\subsection{Taylor expansion}\label{subsecTAYLOR}

The Taylor expansion in $\lambda$ of the cumulants up to order $r$ 
is obtained by using the mixed expansion 
in theorem \ref{th:mixed}, but choosing the order $L$ up to which we develop the loop edges 
to depend on the number of vertices $v$ of the tree $L = \max (0, r - v)$. For $v\ge r+1$ we do not develop any loop edges.
Using the same bounds leading up to 
eq. \eqref{eq:boundseries}, and noting that the scaling with $N$ is always bounded by $N^D$, the rest term is bounded by 
\begin{align*}
 |R_{r}(\lambda,N)| \le & N^{D} 2^{2Dk} k!^{D-1}  \crcr
 & \sum_{v\geq k} \Bigg{[}
 |2\mathcal N_{\mathcal{Q}}\lambda|^{v+q-1} \frac{(2v+k-3+2q)!}{q!   (v-k)!(v+k-1)!} \Bigg{]}_{q= \max(0,r+1-v)}\;,
\end{align*}
hence up to irrelevant overall factors the rest term is bounded by 
\begin{align*}
   &  \sum_{v\ge r+1} |2\mathcal N_{\mathcal{Q}}\lambda|^{v -1} \frac{(2v+k-3)!}{(v-k)!(v+k-1)!} \crcr
 & \quad \quad      +\sum_{v=k}^{r+1}  |2\mathcal N_{\mathcal{Q}}\lambda|^{r} 
 \frac{\left[2v+k-3+2(r+1-v)\right]!}{(r+1-v)!   (v-k)!(v+k-1)!} \; .
\end{align*}
While the first term above is bounded by $|\lambda|^r$ times some constant for $\lambda$ small enough, 
the second one is bounded only as :
\[
 |\lambda|^r \frac{(2r+k-1)!}{(r-k)!} \le (2k-1)! 3^{2r+k-1} \;\;\; r! |\lambda|^r  \; .
\]
Corollary \ref{th:1/N} is straightforward from the previous bounds.

\end{document}